\documentclass{article}
\pdfoutput=1
\usepackage{jheppub}
\usepackage{bm}

\usepackage{tikz}

\newcommand{\roughly}[1]{\mathrel{\raise.3ex\hbox{$#1$\kern-0.85em\lower1ex\hbox{$\sim$}}}}

\newcommand{\abs}[1]{\left|{#1}\right|}

\def\bfA{{\bf A}}

\def\bfB{{\bf B}}

\def\bfe{{\bf e}}
\def\bfE{{\bf E}}
\def\bfJ{{\bf J}}

\def\bfL{{\bf L}}
\def\bfn{{\bf n}}
\def\bfp{{\bf p}}

\def\bfr{{\bf r}}

\def\bfx{{\bf x}}

\def\cA{{\cal A}}
\def\cB{{\cal B}}

\def\cL{{\cal L}}
\def\cM{{\cal M}}
\def\cN{{\cal N}}
\def\cO{{\cal O}}

\def\cR{{\cal R}}
\def\cS{{\cal S}}
\def\cT{{\cal T}}

\def\nn{\nonumber}
\def\({\left(}
\def\){\right)}
\def\[{\left[}
\def\]{\right]}

\newbox\charbox
\newbox\slabox
\def\slsh#1{{      
        \setbox\charbox=\hbox{$#1$}
        \setbox\slabox=\hbox{$/$}
        \dimen\charbox=\ht\slabox
        \advance\dimen\charbox by -\dp\slabox
        \advance\dimen\charbox by -\ht\charbox
        \advance\dimen\charbox by \dp\charbox
        \divide\dimen\charbox by 2
        \raise-\dimen\charbox\hbox to \wd\charbox{\hss/\hss}
        \llap{$#1$}
}}

\def\exd{{\hbox{d}}}

\def\nn{\nonumber}

\def\bea{\begin{eqnarray}}
\def\eea{\end{eqnarray}}
\def\be{\begin{equation}}
\def\ee{\end{equation}}

\def\ssB{{\scriptscriptstyle B}}

\def\QCD{{\scriptscriptstyle QCD}}

\def\pref#1{(\ref{#1})}

\title{Point-Particle Effective Field Theory I: Classical Renormalization and the Inverse-Square Potential}

\author[a,b]{C.P.~Burgess,}
\author[a,b]{Peter Hayman,}
\author[c]{M. Williams}
\author[a,b]{and L\'aszl\'o Zalav\'ari}
\affiliation[a]{Physics \& Astronomy, McMaster University, Hamilton, ON, Canada, L8S 4M1}
\affiliation[b]{Perimeter Institute for Theoretical Physics, Waterloo, Ontario N2L 2Y5, Canada }
\affiliation[c]{Instituut voor Theoretische Fysica, KU Leuven,
Celestijnenlaan 200D,
B-3001 Leuven, Belgium}

\date{\today}

\abstract {Singular potentials (the inverse-square potential, for example) arise in many situations and their quantum treatment leads to well-known ambiguities in choosing boundary conditions for the wave-function at the position of the potential's singularity. These ambiguities are usually resolved by developing a self-adjoint extension of the original problem; a non-unique procedure that leaves undetermined which extension should apply in specific physical systems. We take the guesswork out of this picture by using techniques of effective field theory to derive the required boundary conditions at the origin in terms of the effective point-particle action describing the physics of the source. In this picture ambiguities in boundary conditions boil down to the allowed choices for the source action, but casting them in terms of an action provides a physical criterion for their determination. The resulting extension is self-adjoint if the source action is real (and involves no new degrees of freedom), and not otherwise (as can also happen for reasonable systems). We show how this effective-field picture provides a simple framework for understanding well-known renormalization effects that arise in these systems, including how renormalization-group techniques can resum non-perturbative interactions that often arise, particularly for non-relativistic applications. In particular we argue why the low-energy effective theory tends to produce a universal RG flow of this type and describe how this can lead to the phenomenon of reaction {\em catalysis}, in which physical quantities (like scattering cross sections) can sometimes be surprisingly large compared to the underlying scales of the source in question. We comment in passing on the possible relevance of these observations to the phenomenon of the catalysis of baryon-number violation by scattering from magnetic monopoles. }

\begin{document}

\maketitle
\section{Introduction \& Discussion: point-particle effective field theories}
\label{section:intro}

Effective field theory (EFT) provides an efficient way of exploiting hierarchies of scale when extracting the predictions of physical systems, and the adoption of its methods ever more widely throughout physics has brought the reach of practical calculation to ever more problems over the passage of time. 

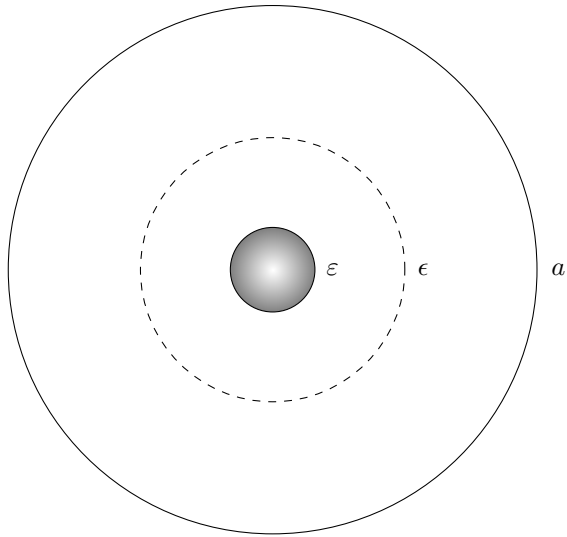
\begin{figure}
	\centering
\begin{tikzpicture}
	\shadedraw[inner color=white, outer color=gray, draw=black] (0,0) circle [radius=16pt];
	\draw (0.8,0) node {$\varepsilon$};
	\draw[dashed] (0,0) circle [radius=50pt];
	\draw (2.0,0) node {$\epsilon$};
	\draw (0,0) circle [radius=100pt];
	\draw (3.8,0) node {$a$};
\end{tikzpicture}
\caption{A schematic of the scales arising when relating near-source boundary conditions to the source action. We denote by $\varepsilon$ the actual UV scale associated with the size of the source (e.g., the size of the proton), which we assume is very small compared to the scale $a$ of physical interest (e.g., the size of an atom). The PPEFT uses the action of the point source to set up boundary conditions on the surface of a Gaussian pillbox of radius $\epsilon$. The precise size of this pillbox is arbitrary, so long as it satisfies $\varepsilon \ll \epsilon \ll a$. We require $\varepsilon \ll \epsilon$ in order to have the first few multipole moments (in our example only the first is considered) dominate the field on the surface of the pillbox, and we require $\epsilon \ll a$ in order to be able to truncate the effective action at the few lowest-dimension terms. The classical RG flow describes how the effective couplings within the PPEFT action must change for different choices of $\epsilon$ in order to keep physical quantities unchanged.} \label{Figcirc}
\end{figure}

So it usually useful to extend its applications to areas where its efficiency of description is not yet completely exploited. In this --- and several companion papers \cite{KG, Dirac, ProtonR} --- we develop EFT tools for use with common problems involving the back-reaction of small `point' sources (PPEFTs), in which the hierarchy of scales to be exploited is the small ratio of the linear size, $\varepsilon$, of the source relative to the size, $a$, of a macroscopic probe. For instance, when applied to atoms (which are more the focus of \cite{KG, Dirac, ProtonR}, rather than this paper) $\varepsilon$ might be the size of the nucleus, regarded as the point source of the nuclear electrostatic potential, while $a$ might be the size of an atomic electron orbit, or the wavelength of a particle that is scattered from the source's gauge or gravitational fields. Our interest is in the limit where $\varepsilon/a \ll 1$ and so for which we can be largely ignorant about the details of the source's underlying structure (see Figure \ref{Figcirc}).

We here lay out the ground-work for the later papers in a simple (and widely studied \cite{EssinGriffiths, Holstein, InvSqPot, Kaplan, singpot, SAExt, SelfAdjExt, Efimov, others}) system: the quantum mechanics of a particle interacting with a point source through an attractive inverse-square potential: $V(r) = - g/r^2$. This problem is well known to be very rich and involve many subtleties to do with its scale-invariance and how this is affected by the boundary conditions that are chosen at $r = 0$ (together with regularization/renormalization issues and dimensional transmutation that follow from these questions). Our treatment here broadly agrees with what is done in the literature, differing mainly in emphasis. We argue though that PPEFTs bring the following two conceptual benefits.

A first benefit of PPEFTs is the clarity they bring to which quantity is being renormalized in these systems. The coupling being renormalized is usually an effective contact coupling within the low-energy point-particle effective action (more about which below). In the simplest situations this coupling turns out to contribute to physics precisely like a delta-function potential would,\footnote{See also \cite{Kaplan}, whose point of view on this is close to our own.} showing that one never really has an inverse-square potential in isolation. Rather, one is {\em obliged} also to include a delta-function coupling whose strength runs --- in the renormalization-group (RG) sense, though purely within the classical approximation --- in a way that depends on the inverse-square coupling $g$. Many otherwise puzzling features of the inverse-square system become mundane once this inevitable presence of the delta-function potential is recognized. For instance, the inverse-square potential is known sometimes to support a single bound state, even when $g$ is small (and sometimes even when it is negative, so the potential is repulsive). It turns out this state is simply the bound state supported by the delta-function, which can remain attractive for the range of $g$ for which the bound state exists.

Using the EFT lens also has a second conceptual benefit. In particular, most treatments of the system agree there is a basic ambiguity to do with how to choose the boundary condition on the wave-function at $r = 0$. This is often phrased as a failure of the Hamiltonian to be self-adjoint \cite{EssinGriffiths, Holstein}, and is usually dealt with by constructing a self-adjoint extension \cite{SAExt, SelfAdjExt}; a process that is not unique (nor appropriate, if the physics of the source should not be unitary, as in \cite{NonUnitary}). The main benefit of thinking in terms of PPEFTs is that it provides an explicit algorithm that directly relates the boundary conditions at $r=0$ to the properties of the effective action describing the point source. This removes the guesswork from the problem, and because it is cast in terms of an action, standard EFT reasoning fixes which boundary conditions should be most important in specific situations (such as at low energies).

Before exploring the inverse-square potential in detail (and some applications) in the next sections, we first pause to flesh out these two claims and to be more explicit about what we mean by a point-particle EFT. 

\subsection{Standard formulations}

There is a regime for which the EFT for point particles is very well understood. When studying the motion of objects like planets or stars within external fields it is common to neglect their internal structure and instead focus completely on their centre-of-mass motion, writing down an effective action for this motion of the form $S = S_\ssB + S_b$, where $S_\ssB$ describes the action of any `bulk' fields --- such as the Maxwell action of electromagnetism for the bulk vector field $A_\mu(x)$, or the Einstein action of gravity for the bulk tensor field $g_{\mu\nu}(x)$ --- while $S_b$ is the `point-particle' action of the centre-of-mass coordinates, $y^\mu$, of the source. 

For instance, the interactions in this point-particle action are often taken to be\footnote{We use units for which $\hbar = 1$.}
\be \label{ptact}
 S_b = \int \exd s \; \cL_b = -  \int  \exd s \; \Bigl[ M \sqrt{- \gamma} - q  A_\mu \, \dot y^\mu  \Bigr] \,,
\ee
where $\dot y^\mu := \exd y^\mu/\exd s$ denotes differentiation along the particle world-line, $y^\mu(s)$, along which $s$ is an arbitrary parameter. Here $\gamma = g_{\mu\nu} \dot y^\mu \dot y^\nu$ denotes the induced world-line metric determinant, and both $g_{\mu\nu}(x)$ and $A_\mu(x)$ are evaluated along the world-line of the source particle of interest: $x^\mu = y^\mu(s)$. The constant parameters $M$ and $q$ ($= +e$ for our main applications) represent the point-particle mass and charge. Variation of $y^\mu$ in this action gives the equations of motion of the particle within a given bulk-field configuration, while variation of $g_{\mu\nu}$ and $A_\mu$ in this action gives the contributions of the particle to the stress-energy and electromagnetic current that respectively source the gravitational and electromagnetic fields. 

PPEFT adds to this framework the effects of particle substructure, which can be incorporated by additional interactions within $S_b$ that involve more fields, more derivatives of the fields, and/or more degrees of freedom (such as spin), in what is essentially a generalized multipole expansion for the sources. Such an expansion has been developed systematically in some situations (such as for slowly moving gravitating objects when computing the orbits and radiation fields of in-spiralling binary systems \cite{binaries}). 

In order to couple the source to the bulk field equations it is useful to re-express $S_b$ as an integral over all of spacetime using a $\delta$-function representation of the form
\be 
\label{brane1}
  S_b = \int \exd s \int \exd^4x \; \cL_b[\phi(x),y(s)] \, \delta^4[x - y(s)] \,,
\ee
where $\phi(x)$ generically denotes any bulk field.\footnote{Although it is tempting to treat the $\delta$-function as independent of the bulk fields this is not in general possible for the metric, since by construction the localizing $\delta$-function is designed to discriminate points according to whether or not they are far enough away to be inside or outside of the source. Assuming it to be metric independent can sometimes lead to inconsistencies with the balance of stress-energy within any microscopic source \cite{DeltaProb}.} When $\cL_b$ is linear in the bulk fields its variation provides the usual inhomogeneous source term for the bulk field field equations. For instance variation of $A_\mu$, which appears linearly in \pref{ptact}, leads to a Maxwell equation of the form
\be
\partial_\mu \Bigl( \sqrt{-g} \; F^{\nu\mu} \Bigr) = q \int \exd s \;  \dot y^\nu(s) \,  \delta^4[x - y(s)] \,,
\ee
which, when specialized to a stationary source, $y^\mu(s) = \{ t(s), 0,0,0 \}$, gives the standard relationship between electric field and the source charge:
\be \label{divE}
  \nabla \cdot \bfE = q \, \delta^3(\bfx) \,.
\ee

The importance of this expression is that its integral over a small spherical gaussian pillbox, $\cS$, of radius $r = \epsilon$, surrounding the point charge, gives the boundary condition that expresses how the integration constants in the bulk fields --- such as the coefficient, $k$, in a spherical solution $\bfE(r) = k \; \bfe_r/r^2$ --- should be related to physical parameters like $q$ appearing in the source action:
\be \label{gauss}
 q = \int_{\cS} \exd^3 \bfx \, q \,\delta^3(\bfx) = \int_{\cS} \exd^3 \bfx \;\nabla \cdot \bfE =  4\pi \epsilon^2 \;\bfe_r \cdot \bfE(r = \epsilon) = 4\pi k  \,.
\ee
The main result for a more general PPEFT is to derive and use the analogue of this expression, which relates directly the near-source boundary conditions in terms of the action describing the properties of the source.

\subsection{Classical renormalization and boundary conditions}

So far so standard. However two related subtleties arise in this approach once terms are examined in $S_b$ that are not simply linear in the bulk fields. One of these is the need that then arises to regularize and renormalize the brane-bulk couplings {\em even at the classical level}. The second is the need to include in this action also the Schr\"dinger field, $\Psi$, and the related change to the boundary conditions of bulk fields that such terms imply. We next briefly summarize both of these issues in turn.

Consider first terms in $S_b$ that are not linear in bulk fields, such as $\cL_b = - c_1 \sqrt{-\gamma} \; (F_{\mu\nu}F^{\mu\nu})^2$ where $c_1$ is a coupling constant. The subtlety is that use of such terms in the above argument involves evaluating the bulk field near the source, and this generically diverges.\footnote{A notable exception is the special case of only one dimension transverse to the source.} For instance a term like $c_1 (\bfE \cdot \bfE)^2$ in $\cL_b$ produces a term of the form $c_1\nabla \cdot \left[ (\bfE \cdot \bfE) \bfE \right] \, \delta^3(\bfx)$ on the right-hand side of \pref{divE}, leading to a similar term in \pref{gauss} whose evaluation involves powers of $\bfE(r = \epsilon) \propto \bfe_r/\epsilon^2$, that diverge for small $\epsilon$. We imagine $\epsilon$ to be related to the physical scales of the problem as sketched in Figure \ref{Figcirc}.

There are several reasons to entertain terms in $S_B$ that are nonlinear in the fields. For the problems studied in this paper we do so because the bulk field we wish to follow is the Schr\"odinger field, $\Psi$, (rather than the electromagnetic field, say) and this field first appears in $S_b$ at quadratic order. But even if we were to focus exclusively on electromagnetic interactions, nonlinear terms are not only allowed, they are usually obligatory for actions describing realistic objects. They arise because they express the dominant implications for long-distance physics of any substructure of the source. 

The divergences that appear in nonlinear terms when bulk fields are evaluated at classical solutions ({\em i.e.} divergences at small $\epsilon$) can be dealt with by appropriately renormalizing source-bulk couplings (couplings like $q$ or $c_1$ in the above example). It is source-bulk couplings (rather than couplings internal to the bulk, say) that renormalize these divergences because the divergences themselves are localized at the source position. Standard power-counting arguments then ensure that renormalization is possible provided all possible interactions are included in the source action that are allowed by the field content and symmetries. This renormalization program has been worked out most explicitly for branes coupled to bulk fields, both for scalars and for gravitational fields\footnote{In the gravitational case some papers \cite{UVCaps} {\em regularize} these divergences, such as by replacing the source by a `thick' brane, but without following through with their renormalization. As argued in \cite{EFTCod2}, such regularization arguments often work in practice - basically because physics far from the sources depends only on a few multipole moments. However, because they do not renormalize they can indicate a dependence on the microphysical scale $\epsilon$ that is misleading.} \cite{GoldbergerWise, deRham, DPR, BvN, BDs, EFTCod2, BDg}, and our presentation here is essentially an adaptation of that developed in \cite{EFTCod2, BDg}. The instance where $S_b$ is quadratic in the fields contains as a special case the well-known phenomenon of renormalization for delta-function potentials \cite{DeltaRun, Jackiw, Holstein}.  

So far as our companion papers are concerned, the most useful result of this paper follows directly from this renormalization story: the renormalization-group (RG) running it implies for the source-bulk couplings. As we shall see, this running turns out to be fairly universal for source actions that stop at quadratic order in the fields (as often dominates at low energies). In some circumstances the running of the source couplings can lead to surprising consequences. In particular, in the presence of an inverse-square potential, $V = - g/r^2$, zero coupling need not be a fixed point of the source-coupling's RG running, and so when $g \ne 0$ it becomes impossible to set the source-bulk couplings to zero at all scales. When this happens they can at best be set to zero at a single scale, say at high energies in the deep ultraviolet (UV). But once this has been done they are free to run, sometimes with unexpected physical consequences at lower energies. 

This RG running of source-bulk couplings sometimes contains surprises. RG invariance generally ensures that physical quantities do not simply depend on the values of the couplings specified at UV scales. Dimensional transmutation instead ensures they are fixed by RG-invariant scales that characterize the flow. (This is reminiscent of how the basic scale of the strong interactions is given by the RG-invariant QCD scale, $\Lambda_\QCD$, rather than the more microscopic scale, $\mu$, at which point the value of the QCD coupling, $g(\mu)$, might be specified.) What is important about these RG-invariant scales is that they can sometimes be much larger than the size of the source being described by the effective theory, and when this is true source-bulk interactions can appear to be surprisingly large. We argue here (and in more detail in \cite{KG, Dirac}) that this fact may partly underlie otherwise-puzzling phenomena like monopole catalysis \cite{monopolecatal, MonopoleReview} of baryon-number violation in Grand Unified Theories (GUTs). 

\subsubsection*{A road map}

Our presentation of these arguments is organized as follows. \S\ref{sec:invpot} examines in detail the running of source couplings that are driven by the non-relativistic quantum mechanics of a bulk Schr\"odinger field interacting with the source through an inverse-square potential in $d$ spatial dimensions. These calculations closely parallel other treatments of inverse-square potentials in the (quite extensive) literature \cite{EssinGriffiths,Holstein, InvSqPot,Kaplan}, for which renormalization effects have been widely studied. Our main new ingredient here is our construction of the boundary condition in terms of the point-particle action. 

\S\ref{sec:apps} then specializes the general arguments of \S\ref{sec:invpot} to scattering in $d=3$ spatial dimensions, with one eye on deriving results of later utility. Along the way we clarify how catalysis (unusually large scattering cross sections) can arise when the RG-invariant scales are much larger than the microscopic size of the source, and apply these results to the special case of non-relativistic $s$-wave scattering of a charged particle (like a proton or electron) from a magnetic monopole. Whether this mechanism actually arises once the low-energy theory is matched to the underlying monopole is a more detailed question that goes beyond the scope of the present paper. 

\section{The Schr\"odinger inverse-square potential in $d$ dimensions}
\label{sec:invpot}

To make the above discussion concrete we now turn to a hoary old saw: the quantum mechanics of a particle interacting with a point source through an inverse-square potential, $V(r) = -g/r^2$, in $d$+1 spacetime dimensions. This system is one that is well studied \cite{EssinGriffiths, Holstein, InvSqPot} and for which the appearance of classical renormalization is well-known. 

The main difference in our presentation is mostly one of emphasis rather than computational difference, with the important exception that we argue there is a systematic way to determine the boundary conditions in terms of the action of the point source. In particular we try to cleanly separate regularization issues (the need to cut off the inverse-square potential in the deep UV regime where $r < \epsilon$) from renormalization issues (the identification of which couplings in the {\em low-energy} theory must be renormalized to cancel the regularization dependence from physical quantities). We do this by interpreting the well-known divergences and dimensional-transmutation that arise for the inverse-square potential in terms of the renormalization of the action of a source situated at the origin, that in the non-relativistic case studied boils down to a delta-function potential.\footnote{This is to be contrasted to a sometimes-articulated alternative picture where what is being renormalized is the value of the cut-off potential, $V(\epsilon)$, in the far-UV regime $r < \epsilon$.} Because renormalization implies the source coupling runs, it {\em must} be present. We argue that its presence helps simplify the understanding of many physical features, such as why bound states sometimes exist in regimes where they are hard to understand purely in terms of the inverse-square potential itself. 

A final reason for re-exploring this system in some detail is that it captures in the simplest context several properties that also arise in more complicated applications, in particular to monopole catalysis of baryon-number violation and novel relativistic effects in Coulomb systems --- that are treated in several companion papers \cite{KG,Dirac, ProtonR}.

\subsection{Schr\"odinger `bulk'}
We start with ordinary quantum mechanics (in $d$ space dimensions) for which $S = S_\ssB + S_b$, with the `bulk' described by the Schr\"odinger action, $S_\ssB$, and where $S_b$ describes microscopic physics of the source situated at the point $\mathbf{r}=0$. In practice we take 
\be
\label{bulkAction}
S_\ssB = \int \exd t \,\exd^d x \left\{ \frac{i}2 \, \Bigl(\Psi^* \partial_t \Psi -  \Psi \, \partial_t \Psi^* \Bigr) - \left[ \frac{1}{2m} \abs{\nabla\Psi}^2 + V(r)\abs{\Psi}^2 \right] \right\} \,,
\ee
where $m$ is the particle mass and
\be
\label{braneAction}
S_b = \int \exd t \cL_b[ \Psi(\bfr=0), \Psi^*(\bfr=0)] = \int \exd t \,\exd^d x \, \cL_b(\Psi, \Psi^*) \, \delta^{(d)}(\bfr) \,,
\ee
with 
\be \label{VLb}
  V(r) = - \frac{g}{r^2} \quad \hbox{and} \quad
  \cL_b = -h \, \Psi^* \Psi 
\ee
used when an explicit form is required. 

The field equation found by varying $\Psi^*$, with the choice \pref{VLb} and specializing to energy eigenstates, for which $\Psi(\bfx,t) = \psi(\bfx) \, e^{-iEt}$, becomes
\be \label{SchE}
  \nabla^2 \psi - U(r)\, \psi = \kappa^2 \, \psi  = -k^2 \, \psi  \,,
\ee
where $\nabla^2 = \nabla_i\nabla^i = \frac 1 {\sqrt{g}}\partial_i(\sqrt{g}\partial^i)$ is the Laplacian for the $d$-dimensional spatial metric, $g_{ij}$, and where 
\be \label{Udel}
  U = 2mV(\mathbf{r}) + 2mh\, \delta^{(d)}(\mathbf{r}) \,,
\ee
and $\kappa^2  = - 2mE = - k^2$. Eqs.~\pref{SchE} and \pref{Udel} show the equivalence of the quadratic source action with a delta-function contribution to the potential. For bound states $E \le 0 \implies \kappa$ is real; for scattering states $E \ge 0 \implies k$ is real. 

In spherical-polar coordinates, the metric for $d$-dimensional Euclidean space can be written $\exd s^2 = g_{ij}\,\exd x^i \exd x^j = \exd r^2 + r^2\hat g_{mn}\,\exd\theta^{m}\,\exd\theta^n$, where $\hat g_{mn}$ is the metric on the ($d-1$)-sphere, parameterized entirely by the periodic coordinates $\theta^1, \dots, \theta^{d - 1}$. In terms of this metric, the Schr\"odinger equation is separable into a radial piece and an angular piece:
\be \label{SchESep}
\frac{1}{\sqrt{g}}\partial_i\left(\sqrt{g}\,\partial^i \psi\right) - U(r)\psi = \frac{1}{r^{d-1}}\frac{\exd}{\exd r}\left(r^{d - 1}\frac{\exd}{\exd r}\psi\right) + \frac{1}{r^2}\Delta \psi - U(r) = \kappa^2 \psi\,,
\ee
where $g$ is the metric determinant $g = r^{2d - 2}\hat g$, and $\Delta$ is the Laplacian for $\hat g_{mn}$ on the ($d-1$)-sphere. 

Clearly the equation is separable and we can write
\be
\psi(r,\theta^1,\theta^2, \dots, \theta^{d-1}) = \sum_{\omega} \psi_{\omega}(r) \, Y_{\omega}(\theta^1,\theta^2, \dots, \theta^{d - 1}) \,,
\ee
where $\omega$ is one or more parameters associated with the $d$-dimensional spherical harmonics $Y_\omega$. (For instance $\omega = \ell$ is an integer when $d=2$ and $\omega = \{\ell, \ell_z \}$ if $d=3$.) If we define the eigenvalue of the ($d-1$)-dimensional spherical laplacian by $-\varpi$ (e.g., this is the usual $-\ell(\ell + 1)$ in three dimensions, while it is $-\ell^2$ in two dimensions, and of course 0 in one dimension), then the Schr\"odinger equation reduces to the following radial equation for $\mathbf{r} \not = 0$:
\be \label{SchEr}
\frac{1}{r^{d -1}}  \frac{\exd }{\exd r} \left( r^{d - 1} \frac{\exd \psi_{\omega}}{\exd r} \right) - \left[ \frac{\varpi - \alpha }{r^2} \right] \psi_{\omega} 
 = \kappa^2 \, \psi_{\omega}  \,,
\ee
where $\alpha = 2m g$. The solutions to this equation are $(\kappa r)^{-(d-2)/2}I_{l+(d-2)/2}$ and $(\kappa r)^{-(d-2)/2}K_{l + (d - 2)/2}$, where $I_{l + (d-2)/2}(\kappa r)$ and $K_{l + (d - 2)/2}(\kappa r)$ are the modified Bessel functions of order $l + (d - 2)/2$, where $\l$ is defined by $2l + d - 2 = \zeta$, and $\zeta := \sqrt{(d - 2)^2 + 4(\varpi - \alpha)}$.

\subsection{Near-source boundary condition}
\label{sec:BC}
For small $r$ the radial solutions asymptote to become proportional to $r^{l}$ and $r^{-l - (d - 2)}$. For $d \geq 3$, these can {\em both} diverge as $r \to 0$ for some $\alpha$, such as for $\varpi = 0$ and $0 < \alpha < (d - 2)^2/4$ (indeed, the two functions $|\psi_\pm|$ both diverge in precisely the same way --- as $r^{-(d - 2)/2}$ --- once $\zeta$ vanishes or becomes imaginary, such as for $\varpi = 0$ and $\alpha \ge (d - 2)/4$). This shows that boundedness of the solutions as $r \to 0$ {\em cannot} be the right boundary condition to use at the origin. 

It is perhaps not a big surprise that boundedness is not the right criterion because there are many examples where fields diverge at the positions of point sources, such as does the Coulomb potential at the position of a point charge. And once this point is conceded one must recognize that boundedness also cannot be the right criterion to use in general, and in particular it might not be appropriate to discard solutions in situations for which one solution is bounded and the other one isn't (such as in two dimensions when $\varpi = \ell^2 = 1$, and $\alpha = 0$ so that the solutions go as $r^1$ and $r^{-1}$), given the presence of a source. 

What boundary condition should be imposed instead? A weaker criterion at the origin imposes the normalizability of $\psi$, which asks $\int \exd r \, r^{(d - 1)} |\psi|^2$ to converge in the regime $r \to 0$. This condition excludes solutions that diverge faster than $\psi \sim r^{-d/2}$. When $\alpha < d(d-4)/4 + \varpi$, we find $\zeta > 2$, and so $-2l - 2(d-2) + (d - 1) = -\zeta - (d - 2) + (d - 1) = -\zeta + 1 < -1$, hence the solution that goes as $r^{-l -(d-2)}$ is excluded. 

The case of most interest in what follows is when $d(d-4)/4 < \alpha < d(d-4)/4 + \varpi_1$, where $\varpi_1$ denotes the first non-zero eigenvalue of the angular laplacian\footnote{Of course in one dimension, $\varpi$ does not strictly exist, since there is no angular Laplacian. In that case, we define $\varpi_1$ as $\infty$, since the lack of higher angular momentum state means there is no upper bound on $\alpha$.}, as this is the region for which normalizability excludes the solutions going like $r^{-l - (d-2)}$ for all but the $\varpi = 0$ $s$-wave states (e.g.~in three dimensions, this is the range $-3/4 \le \alpha < 5/4$). Normalizability is insufficient in itself in this case to determine the boundary condition for $s$-wave states. It is for these that we argue the correct condition instead is given by the properties of the source action, $S_b$, as indicated below. Physically this occurs because the inverse-square potential draws the $s$-wave wavefunction sufficiently towards the origin that the net flow of probability there cannot be determined without knowing more precisely how the particle interacts with the source located at $r = 0$. For small enough $\alpha$ the centrifugal barrier is strong enough to keep this from happening for any nonzero $\varpi$.

The choice of boundary conditions as $r \to 0$ also bears on whether or not the inverse-square Hamiltonian is self-adjoint, and this is the way the need to choose boundary conditions is usually framed. Although usually not incorrect,\footnote{Unless the physics of the source {\em does} allow it to be a sink --- or source --- of probability, such as in situations like those described in \cite{NonUnitary}, for example.} demanding boundary conditions ensure self-adjointness ({\em i.e.} finding a self-adjoint extension) typically does not determine them uniquely. The advantage of casting the boundary condition directly in terms of the source action is that it makes explicit the connection between any non-uniqueness and the choices available for the physics of the source. When the source action is real the resulting boundary conditions ensure no loss of probability, as we show in a specific example below.

The implications of the source action, $S_b$, for the boundary condition is obtained\footnote{Strictly speaking the reasoning presented here is only true for delta-function interactions in the absence of inverse-square potentials, because the singularity of the $-g/r^2$ potential at $r = 0$ undermines the argument that only the derivatives and delta function can contribute when integrating the equations of motion over a small pillbox. A better derivation that also applies when inverse potentials are present is given in Appendix \ref{AppBC}, with the bonus that its formulation also provides a clearer picture of what the RG equations physically mean.} by integrating \pref{SchE} over an infinitesimal sphere, $\cS$, of radius $0 \le r \le \epsilon$ around $\bfr = 0$ (and ignoring the $-g/r^2$ potential, as discussed in Appendix \ref{AppBC}). For the simple quadratic action considered here the result is the same as expected for a delta-function potential where continuity of $\psi$ implies the result gets only contributions from the delta function and from the integral of the second derivative, leading to the result
\be
\label{preSphbc}
\lambda \, \psi(0) = \int_\cS \exd^d x \, \nabla^2 \psi =  \int_{\partial \cS} \exd^{d -1} x \, \bfn \cdot  \nabla \psi =   \int \exd\Omega_{d - 1} \, \left( r^{d - 1} \,\frac{\partial \psi}{\partial r} \right)_{r = \epsilon} = \Omega_{d -1} \epsilon^{d -1} \left( \frac{\partial \psi}{\partial r} \right)_{r = \epsilon} \,,
\ee
where $\lambda = 2mh$, $\bfn \cdot \exd \bfr = \exd r$ is the outward-pointing radial unit vector, $\exd\Omega_n = \sqrt{\hat g}\, \exd \theta^1 \exd \theta^2\dots\exd\theta^{n}$ is the volume element on the surface of the unit $n$-sphere, and $\Omega_{d -1}$ is the corresponding volume. The last equality assumes a spherically symmetric source and that $\epsilon$ is small enough that $\psi$ is also spherically symmetric to good approximation. 

The required boundary condition at the origin then is 
\be \label{sphbc}
\left[\Omega_{d -1} r^{d -1} \,  \frac{ \partial}{\partial r} \ln \psi \right]_{r=\epsilon}  =  \lambda\,,
\ee
which uses the definition $\psi(0) := \psi(r=\epsilon)$. Note that in one dimension, $r$ can take negative values, and so the integral \eqref{preSphbc} evaluates to
\be \label{sphbcD1}
\left[\frac{ \partial}{\partial r} \ln \psi \right]_{r=-\epsilon}^{r = \epsilon}  =  \lambda \qquad (d = 1).
\ee
But since the Schr\"odinger equation is invariant under $r \to - r$, the normalizable solution $\psi_<$ for $x<0$ is equal to the normalizable solution $\psi_>$ for $x>0$ evaluated at $-x$, i.e. $\psi_<(x) = \psi_>(-x)$. This means $\partial_x \ln \psi \vert_{r = -\epsilon} = -\partial_x \ln \psi \vert_{r = \epsilon}$, and so if we define $\Omega_{0} := 2$, then the boundary condition \eqref{sphbc} is valid for all dimensions $d \ge 1$.

As we see later, this boundary condition implies physical quantities depend on $\lambda$, and (at face value) also on $\epsilon$. The main point in what follows is that the dependence of all physical quantities on $\epsilon$ can be absorbed into an appropriate renormalization of the parameter $\lambda$. After this is done all explicit $\epsilon$-dependence becomes cancelled by the implicit $\epsilon$-dependence in $\lambda$ implied by the boundary condition \pref{sphbc} or \pref{sphbcD1}. Before pursuing this further we pause briefly to ask whether or not this boundary condition is unitary. 

\subsection{Source action vs self-adjoint extension}

The inference of the boundary condition at $r = 0$ from the source action, $S_b$, presented above encodes how the source back-reacts onto its environment over distances much larger than the size of the source itself. Does this boundary condition provide a self-adjoint extension \cite{SAExt, SelfAdjExt} in the sense of conserving probability at the source?
  
To see how this works we use the boundary condition to compute whether the region $r < \abs{\epsilon}$ is a source or sink of probability using the radial probability flux, 
\be
J = \Omega_{d-1} r^{d -1} \, \bfn \cdot \bfJ = \frac{\Omega_{d-1} r^{d -1}}{2m} \Bigl(  \Psi \partial_r \Psi^* - \Psi^* \partial_r \Psi  \Bigr) \,.
\ee
Evaluating with energy eigenstates gives
\bea
J(\epsilon) &=& \frac{\Omega_{d-1} \epsilon^{(d -1)}}{2m} \Bigl[ \psi(\epsilon)\partial_r \psi^*(\epsilon) - \psi^*(\epsilon) \partial_r \psi(\epsilon)\Bigr]\nn\\
&=&  (h^* - h) \psi^*(\epsilon) \psi(\epsilon) \qquad (d > 1)\,,
\eea
and
\be
J(\epsilon) = J(-\epsilon) + (h^* - h) \psi^*(\epsilon) \psi(\epsilon) \qquad (d = 1)\,.
\ee
This states something reasonable: probability is conserved at the source provided either its action is real (ie $h^* = h$) and/or there is no probability of finding the particle at the source (ie $\psi^*(0) \psi(0) = 0$). Alternatively, $h$ can instead be chosen to be complex should a non-unitary boundary condition be desired.\\

\subsection{Bound States}
We next seek bound-state solutions to the bulk equations, using their small-$r$ form to determine how $\lambda$ runs with $\epsilon$. (Clearly this running also could be determined using scattering solutions rather than bound states. As we see explicitly below the results are the same.) Although these solutions can be written in terms of modified Bessel functions, we here instead analyze them in terms of confluent hypergeometric functions since this generalizes more easily to other applications (such as those in \cite{KG,Dirac}). To this end we write the radial equation \pref{SchEr} in the form
\be \label{SchEgk}
r^2 \frac{\exd^2 \psi}{\exd r^2} + (d - 1)r \, \frac{\exd \psi}{\exd r} + \left( v - \kappa^2 r^2 \right) \psi
 = 0\,,
\ee
where $v = \alpha - \varpi$. This is put into the confluent hypergeometric form if we define $\psi(z) = z^l \, e^{-z/2} u(z)$, for $z = 2\kappa r$ and constant $l$ satisfying $l(l + d - 2) + v = 0$, and so\footnote{Choosing the other root for $l$ just exchanges the roles of the two independent solutions encountered below, so does not introduce any new alternatives.} given by $l = \frac 12 ( 2 - d + \zeta)$. In this case the two linearly independent radial profiles can be written 
\begin{equation}
	\psi_\pm = (2\kappa r)^{\frac 12\left(2 - d\, \pm\, \zeta\right)}e^{-\kappa r}\cM\left[\frac 12\left(1 \pm \zeta\right), 1 \pm \zeta; 2\kappa r\right].
\end{equation}
where $\cM(a,b;z) = 1 + (a/b) z + \cdots$ is the usual confluent hypergeometric function and 
\be
\zeta := \sqrt{(d - 2)^2 + 4(\varpi - \alpha)} = 2l+ d - 2\,,
\ee
as before.

For bound states we seek solutions normalizable at infinity, and the large-$z$ asymptotic expansion of hypergeometric functions shows this leads to the following combination of solutions (with arbitrary normalization constant $C$)
\be \label{pmform}
 \psi_\infty(r) = C \left[  \frac{\Gamma(-\zeta)}{\Gamma\left[\frac12\left(1 - \zeta\right)\right]} \; \psi_+(r) +
 \frac{\Gamma(\zeta)}{\Gamma\left[\frac12\left(1 + \zeta\right)\right]} \; \psi_-(r) \right]  \,,
\ee
which shows that integer $\zeta$ can be problematic (and so is obtained by a limiting procedure if it arises).

On the other hand the solutions $\psi_\pm(r)$ for small $r = \epsilon$ behave as
\be
\psi_\pm(\epsilon) = (2\kappa \epsilon)^{\frac12(2 - d \pm \zeta)} \left[ 1  + \cO(\epsilon^2) \right] \,,
\ee 
and so $\psi_\pm$ is only normalizable at $r = 0$ when $2 \pm \zeta > 0$. Adopting the convention that $\zeta \ge 0$ when real, we see that $\psi_+$ is always normalizable at $r = 0$ but $\psi_-$ is only normalizable there when $\zeta < 2$. Keeping in mind that $\zeta = \sqrt{(d - 2)^2 + 4(\varpi - \alpha)}$ we see (in agreement with the discussion above) that $\psi_-$ can be discarded for all $\varpi \ne 0$ whenever $d(d-4)/4 \le \alpha < d(d - 4)/4 + \varpi_1$ for the first non-zero $\varpi := \varpi_1$, and the inability to choose a value for $\kappa$ in \pref{pmform} to ensure this implies no bound states exist in this case for nonzero $\varpi$.

For $d(d-4)/4 \le \alpha < d(d - 4)/4 + \varpi_1$, either $\zeta < 2$ or $\zeta$ is imaginary when $\varpi = 0$, and so both solutions are normalizable. In this case we instead use the boundary condition at the origin given in \pref{sphbc} which, when using the small-$r$ expansion,
\be
\left(  \frac{\partial \psi_\pm}{\partial r} \right)_{r=\epsilon} = \kappa \left( 2 - d \pm \zeta \right) (2\kappa \epsilon)^{\frac12(-d\pm\zeta)} 
+ \cdots \,,
\ee
becomes
\bea \label{bceps}
\lambda &=& \Omega_{d-1} \epsilon^{d - 1} \left( \frac{\partial }{\partial r} \,\ln \psi \right)_{r=\epsilon} \notag \\
&=& \Omega_{d-1}  \kappa \epsilon^{d -1}\left[ \frac{C_+ \left( 2- d + \zeta \right) (2\kappa \epsilon)^{\frac12(-d+\zeta)} + C_- \left( 2 - d - \zeta \right) (2\kappa \epsilon)^{\frac12(-d-\zeta)}}{ C_+  (2\kappa \epsilon)^{\frac12(2-d+\zeta)} + C_-   (2\kappa \epsilon)^{\frac12(2-d-\zeta)}} \right] \nn\\
&=&\Omega_{d-1}  \frac{\epsilon^{d - 2}}{2} \left[ \frac{ \left( 2-d + \zeta \right)+ R \left( 2-d - \zeta \right)}{ 1+ R} \right] \nn\\
 &=& -\Omega_{d-1}  \frac{\epsilon^{d - 2}}{2} \left[ d - 2 + \zeta \left( \frac{R-1}{R+1}\right)  \right]  \,,
\eea
where
\be \label{Rexp}
 R := \left( \frac{C_-}{C_+} \right) \, (2\kappa \epsilon)^{-\zeta} \,.
\ee
Notice $R=0$ when $C_- = 0$, which for sufficiently small $\alpha$ applies for all nonzero choices of $\varpi$.

To use this equation we rewrite it as
\be \label{bceps2}
\hat \lambda := \frac{2\lambda}{\Omega_{d-1} \epsilon^{d - 2}}  + d - 2 =   \zeta \left( \frac{1-R}{1+R} \right) \,,
\ee
where the first equality defines $\hat\lambda$. Notice that the delta-function potential is repulsive when $\lambda > 0$ and so $\hat \lambda > d-2$, while attractive $\delta$-potentials ($\lambda < 0$) imply $\hat\lambda < d-2$. 

Once values for $\zeta$ and $\hat\lambda$ are given, eq.~\pref{bceps2} is used by solving it for $R$ in terms of the given $\hat\lambda$ and $\zeta$:
\be \label{Rvslamzet}
 R(\hat\lambda) = \frac{\zeta - \hat \lambda}{\zeta + \hat \lambda} \,.
\ee
This is positive if $|\hat \lambda| \le \zeta$ and negative otherwise. Equating this to \pref{Rexp} either allows the determination of $C_-/C_+$ (in the case of scattering, more about which below) or the value of $\kappa$ if $C_-/C_+$ is already determined as in the case of a bound state, where \pref{pmform} fixes $C_-/C_+$ and so implies
\be
\label{BigR}
  R =  \frac{\Gamma(\zeta)\Gamma\left[\frac12\left(1 - \zeta\right)\right]}{\Gamma(-\zeta)\Gamma\left[\frac12\left(1 + \zeta\right)\right]} \, (2\kappa \epsilon)^{-\zeta} =  \frac{\Gamma\left(\frac12\zeta\right) }{\Gamma\left(-\frac12\zeta\right)} \, \left(\frac{\kappa \epsilon}2 \right)^{-\zeta}  \,,
\ee
which uses Euler's duplication formula. Since $\zeta$ is a square root, it will either be entirely real or imaginary. Given that, \eqref{BigR} can be written $R = -e^{\beta}$, with 
\be
\label{betaIm}
\beta = \ln\;\left[-\frac{\Gamma(\frac 12 \zeta)}{\Gamma(-\frac 12\zeta)}\right] - \zeta\ln\left(\frac{\kappa\epsilon} 2\right).
\ee
With this, we can write \eqref{bceps2} cleanly as
\be
\label{lambdaBeta}
\hat\lambda = -\zeta\coth\left(\frac \beta 2\right).
\ee
It should be noted that when $\zeta = i\xi$ is purely imaginary, $\beta = i\nu$ is also purely imaginary, and the hyperbolic cotangent becomes a regular cotangent. That is,
\be
\label{lambdaBetaIm}
\hat\lambda = -\xi\cot\left(\frac \nu 2\right). 
\ee

Notice that \eqref{bceps} either diverges or vanishes as $\epsilon \to 0$, and this is where the renormalization story comes in. We must choose $\lambda$ also to diverge or vanish as $\epsilon \to 0$ in such a way as to ensure what remains is a finite, sensible $\epsilon$-independent expression for $\kappa$ (and so also for the bound state energy $E$).

\subsection{Renormalization}
Again focusing on the case $d(d-4)/4 \le \alpha < d(d - 4)/4 + \varpi_1$, we first consider the case where $\zeta_s := \zeta(\varpi = 0) = i\xi_s$ is imaginary, and so $(d - 2)^2/4 \le \alpha < d(d-4)/4 + \varpi_1$. To determine in this case how $\lambda$ must depend on $\epsilon$ in order to renormalize any divergences as $\epsilon \to 0$, we use that the energy $\kappa$ cannot depend on $\epsilon$ and differentiate \eqref{lambdaBetaIm} using $\epsilon\partial\nu/\partial\epsilon = -\xi_s$ to get
\be
\label{run1Im}
\epsilon\frac{\partial}{\partial \epsilon}\left(\frac{\hat\lambda}{\xi_s}\right) = - \frac{\xi_s}{2\sin^2(\nu/2)} = -\frac{\xi_s}{2}\left[1 + \cot^2\left(\frac \nu 2\right)\right] = -\frac {\xi_s} 2\left[1 + \left(\frac{\hat\lambda} {\xi_s}\right)^2\right].
\ee
This can be integrated to find $\hat\lambda(\epsilon)$ giving
\be
\label{run2Im}
\frac{\hat\lambda(\epsilon)}{\xi} = \frac{\hat\lambda_0 - \xi\tan\left[\frac{\xi}{2}\ln(\epsilon/\epsilon_0)\right]}{\xi + \hat\lambda_0\tan\left[\frac{\xi}{2}\ln(\epsilon/\epsilon_0)\right]},
\ee
where $\hat\lambda_0 := \hat\lambda(\epsilon_0)$ is used to choose the integration constant.

Inserting this into \eqref{lambdaBetaIm} gives $\kappa$ in a manifestly $\epsilon$-independent way, with a simple calculation showing that $\kappa$ is again given by \eqref{lambdaBetaIm}, but with $\hat\lambda \to \hat\lambda_0$ and $\nu \to \nu_0 := \nu(\epsilon_0)$:
\be
\label{run3Im}
\hat\lambda_0 = -\xi\cot\left(\frac {\nu_0} 2\right). 
\ee
For instance, if the arbitrary scale $\epsilon_0$ is chosen such that $\hat\lambda_0 = 0$ (i.e., such that $\lambda(\epsilon_0) = (2-d)\Omega_{d-1} \epsilon_0^{d-2}/2$), then if $\xi \neq 0$, we must have $\cot(\nu_0/2) = 0$, and so $\kappa$ is given explicitly in terms of $\epsilon_0$ by
\be
\label{quantIm}
\nu_0(\kappa) = (2n + 1)\pi = \hbox{arg}\;\left[-\frac{\Gamma(\frac 12 i\xi_s)}{\Gamma(-i\frac12 \xi_s)}\right] - \xi\ln\left(\frac{\kappa\epsilon_0} 2\right).
\ee
Solving then gives
\be
\label{energyIm}
\kappa = \frac 2 {\epsilon_0}e^{[\theta(\xi) - (2n + 1)\pi ]/\xi},
\ee
where $e^{i\theta} := -\Gamma(i\xi_s/2)/\Gamma(-i\xi_s/2)$.

This indicates an infinite number of states
\be
\label{energyIm2}
E_n = -\frac{\kappa^2}{2m} = -\frac 2 {m\epsilon^2_0}e^{2[\theta(\xi) - (2n + 1)\pi]/\xi},
\ee
where $n = 0,\pm1,\pm2,\cdots$ up until $n$ is so large that $|E_n|$ is greater than the UV cutoff of the EFT in which the source size is not resolved (allowing it to contribute simply a delta function). As is often observed, dimensional transmutation is at work here with the appearance in $E_n$ of the scale $\epsilon_0$. The above discussion shows that it is the value of the dimensionless coupling $\hat \lambda$ that is traded for the scale $\epsilon_0$ (as opposed to the dimensionless coupling, $\alpha$, say). 

This is the RG limit-cycle regime that has been extensively studied, in particular within the context of the Efimov effect \cite{Efimov}.

\begin{figure}[h]
\begin{center}
\includegraphics[width=90mm,height=60mm]{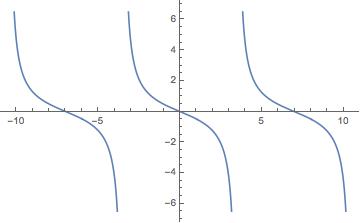} 
\caption{Plot of the RG flow of $\hat\lambda/\xi_s$  vs $\ln \epsilon/\epsilon_\star$ where $\hat \lambda = (\lambda/2\pi\epsilon)+1$ and $(d-2)^2/4 < \alpha < d(d-4)/4 + \varpi_1$ is chosen so that $\zeta_s = i \xi_s$ is imaginary. $\hat\lambda$ strictly decreases as $\epsilon$ grows, and the parameter $\epsilon_\star$ is chosen such that $\hat\lambda(\epsilon_\star) = 0$.} \label{fig:RGflow2} 
\end{center}
\end{figure}

\subsubsection*{The limit $\alpha \to (d-2)^2/4$}

The case where $\alpha \to (d-2)^2/4$ (and so $\zeta_s = i\xi_s \to 0$) is instructive for two reasons. First, the tower of bound states collapses to a single bound state, and the reason for the one remaining state is clear: in this limit the inverse-square potential is not deep enough to support a bound state and so the  lone bound state is the one supported by the delta-function potential. Second, dimensional transmutation for this state can allow an exponential suppression of the binding energy of this state relative to the typical (microscopic) scale set by the delta-function itself, and this is what allows the bound state to have low-enough energy to be reliably described purely within the low-energy theory. 

Recalling the small-$\nu$ form $\Gamma(1 \pm i\xi_s/2) = 1 \mp i \gamma \xi_s/2 + \cO(\xi_s^2)$, and so $e^{i\theta(\xi_s)} = \Gamma(1+i\xi_s/2)/\Gamma(1-i\xi_s/2) \simeq 1 -i\gamma\xi_s + \cO(\xi_s^2)$, with $\gamma \simeq 0.5772\cdots$ the Euler-Mascheroni constant, we have $\theta(\xi_s) = -\gamma\xi_s + \cO(\xi_s^2)$ as $\xi_s \to 0$. Therefore 
\be
 \lim_{\xi_s \to 0} \frac{\nu_0}{\xi_s} = - \left[ \gamma + \ln \left( \frac{\kappa \epsilon_0}2 \right) \right] \,,
\ee
and so in this limit \eqref{run3Im} becomes
\be \label{bcreg0nu0}
\hat\lambda_0 = -\lim_{\xi_s \to 0} \xi_s \cot \frac{\nu_0}2 =  -\lim_{\xi_s \to 0} \frac{2\xi_s}{\nu_0} = \frac{2}{\gamma + \ln(\kappa \epsilon_0/2)}\,,
\ee
and so the bound-state condition becomes
\be \label{kappanu0}
 \kappa = \frac{2}{\epsilon_0} \; \exp \left[ - \gamma + \frac{2}{\hat\lambda_0} \right] \,.
\ee

As $\xi_s \to 0$ the running of $\lambda$ is most easily found by returning to the RG equation \pref{run1Im} and integrating it from scratch, leading to
\be \label{RGnu0}
 \epsilon \, \frac{\partial \hat\lambda}{\partial \epsilon} =  - \frac{\hat\lambda^2}2 \,,
\ee
which integrates to
\be
\frac 1 {\hat\lambda(\epsilon)} = \frac 1 {\hat\lambda_0} + \frac{1}{2}\ln(\epsilon/\epsilon_0).
\ee
Again this is precisely what is required to make $\kappa$ determined from \pref{kappanu0} independent of $\epsilon$. 

A convenient way to write this is in terms of the `QCD' scale, $\epsilon_\star$, defined as the scale where $\hat\lambda \to \infty$ becomes strong. In terms of $\epsilon_\star$ \pref{kappanu0} becomes
\be
 \kappa = \frac{2}{\epsilon_\star} \; e^{-\gamma} \,,
\ee
which shows that it is $\epsilon_\star$ that sets the scale of the lone bound state in this limit. This last result is only really interesting if $\epsilon_\star \gg \epsilon_0$, since this allows a hierarchy between the bound state and the scale of the `brane' source.  We see that $\hat\lambda \to +\infty$ for $\epsilon \to \epsilon_\star$ satisfying 
\be
  \epsilon_\star  = \epsilon_0 \, \exp\left[- 2/\hat\lambda_0\right] \,. 
\ee

These last two equations show that the bound state scale $\epsilon_\star$ is only much greater than the microscopic scale $\epsilon_0$ when $\hat \lambda_0$ is negative and very small. When this is not true the bound state is not macroscopic and so its existence cannot reliably be inferred purely within the low-energy effective theory being used. It is in this way that we see why the existence of the bound state is in this case only reliably inferred when $\lambda_0$ is slightly below the value $1/\epsilon_0$. 

\subsubsection*{The Case $\alpha < (d - 2)^2/4$}

Although the inverse-square potential does not support bound states when $\alpha < (d - 2)^2/4$, the delta-function potential can continue to do so, as we now show. In this regime, the bound-state condition is given by \eqref{lambdaBeta}, which we repeat here
\begin{equation*}
	\hat\lambda = -\zeta_s\coth\left(\frac \beta 2\right). \tag{\ref{lambdaBeta}}
\end{equation*}
This admits no solutions if $\abs{\hat\lambda} < \zeta_s$ and admits one solution when $\abs{\hat\lambda} > \zeta_s$. 

The running of $\hat\lambda$ is again found by differentiating the quantization condition with respect to $\epsilon$, holding $\kappa$ fixed. When $\zeta_s$ is real we use $\epsilon\, \exd \beta/\exd \epsilon = -\zeta_s$ and $\exd \kappa/\exd \epsilon =  \exd \zeta/\exd \epsilon = 0$, to find
\be
 \epsilon \, \frac{\exd}{\exd \epsilon} \left( \frac{\hat\lambda}{\zeta_s} \right) = - \frac{\zeta_s}{2 \sinh^2 ({\beta}/2)} = \frac{\zeta_s}{2} \left[ 1 - \coth^2 \frac{\beta}2  \right] = \frac{\zeta_s}2 \left[1- \left( \frac{ \hat \lambda}{\zeta_s} \right)^2  \right] \,.
\ee
This integrates to
\be \label{lambdaeps}
 \frac{\hat \lambda(\epsilon)}{\zeta_s} = \frac{ (\hat \lambda_0/\zeta_s) + \tanh\left[ \frac12 \, \zeta_s \ln(\epsilon/\epsilon_0) \right]}{1 + (\hat \lambda_0/\zeta_s) \, \tanh\left[ \frac12 \, \zeta_s \ln(\epsilon/\epsilon_0)\right]} \,,
\ee
where $\hat\lambda_0 := \hat \lambda(\epsilon_0)$ is again used to choose the integration constant.

\begin{figure}[h]
\begin{center}
\includegraphics[width=90mm,height=60mm]{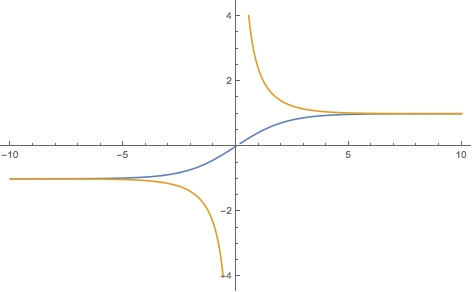} 
\caption{Plot of the RG flow of $\hat\lambda/\zeta_s$ vs $\ln \epsilon/\epsilon_\star$ where $\hat \lambda = (\lambda/2\pi \epsilon) +1$ and with $d(d-4)/4 < \alpha < (d-2)^2/4$ chosen so that $\zeta_s$ is real. A representative of each of the two RG-invariant classes of flows is shown, and $\epsilon_\star$ is chosen as the place where $\hat \lambda = 0$ or $\hat \lambda \to \infty$, depending on which class of flows is of interest.} \label{fig:RGflow} 
\end{center}
\end{figure}

It is important to note that this RG flow now has fixed points\footnote{Ref.~\cite{Kaplan} uses the disappearance of these fixed points as $\alpha \to \frac14$ as an archetype of how fixed-point coalescence can lead to universal features.} at 
\be \label{fixedpt}
   \hat \lambda_\pm := \hat \lambda[\ln (\epsilon/\epsilon_0) \to \pm\infty]  = \pm \zeta_s\,, 
\ee
with the flow towards the infrared (large $\epsilon$) going from $\hat\lambda_-$ to $\hat\lambda_+$. As shown in Fig.~\ref{fig:RGflow}, this evolution supports the following two disjoint classes of flow, according to the size of $|\hat\lambda|/\zeta_s$ throughout the flow:
\begin{itemize}
\item One class satisfies $|\hat\lambda| \le \zeta_s$ 
for all $\epsilon$. RG flows within this class climb monotonically from $\hat\lambda_-$ to $\hat\lambda_+$ as $\epsilon$ increases, ensuring that there always exists a scale, $\epsilon_0$, for which $\hat \lambda(\epsilon_0) = 0$. Since bound states require $|\hat \lambda| \ge \zeta_s$ 
systems that flow along these trajectories never have delta-function couplings that can support any bound states.  
\item The second class of flows satisfies $|\hat\lambda| \ge \zeta_s$ 
for all $\epsilon$. RG flows within this class initially drop monotonically through negative values of $\hat \lambda$ from $\hat\lambda_-$ until $\hat\lambda \to - \infty$ as $\epsilon \to \epsilon_\star$ from below. For $\epsilon > \epsilon_\star$ the coupling $\hat \lambda$ falls from $\hat \lambda \to + \infty$ as $\epsilon \to \epsilon_\star$ from above, and falls monotonically with increasing $\epsilon$ until eventually approaching $\hat\lambda_+$ as $\epsilon \to \infty$. Systems with these RG flows satisfy $|\hat \lambda| \ge \zeta_s$ 
and so can support a bound state. We shall see that the bound state arises with characteristic scale $\epsilon_\star$, and this is only macroscopic when $\hat \lambda$ is negative and very small, just as was found above for $\alpha = (d-2)^2/4$. 
\end{itemize}

Focusing on the case $|\hat \lambda| > \zeta_s$, the bound-state condition is efficiently given in terms of the scale $\epsilon_\star$. Setting $\hat\lambda_\star = \hat\lambda(\epsilon_\star) \to \infty$ in the bound state condition implies $\coth (\beta/2) \to \infty$ and so $\beta \to 0$, implying $R = -1$ and so
\be \label{kappavsepsstar}
  \kappa = \frac{2}{\epsilon_\star} \left[ -\frac{ \Gamma(\frac 12\zeta_s)}{\Gamma(-\frac 12\zeta_s)} \right]^{1/\zeta_s} 
 \,,
\ee
where $\epsilon_\star$ is given in terms of $\epsilon$ and $\hat\lambda(\epsilon)$ by
\be
\label{tanhStar}
  \tanh \left[ \frac 12\zeta_s\ln \left( \frac{\epsilon_\star}{\epsilon} \right) \right] = - \; \frac{\zeta_s}{\hat \lambda } \,.
\ee
For $\zeta_s$ positive and $\epsilon_\star \gg \epsilon$ the left-hand side is just a hair smaller than $+1$ and this requires $\hat \lambda/\zeta_s= -(1+  \delta)$ with $0 < \delta \ll 1$. Using in this regime the asymptotic expression $\tanh x=(1 - e^{-2x})/(1 + e^{-2x}) \simeq 1 - 2 e^{-2x}$ for large $x$ then leads to the approximate expression  
\be
   \epsilon_\star \simeq \epsilon \left( \frac{2}{\delta} \right)^{1/\zeta_s} \,.
\ee
These show how the formulae for a bound state with macroscopic size become extended into the regime where $\alpha < (d-2)^2/4$ for $\hat \lambda$ sufficiently close to $- \zeta_s$. In this regime the underlying delta-function potential (whose strength is $\lambda$) is sufficiently attractive to give rise to this bound state, and this is its physical origin. 

The RG running of $\lambda$ can sometimes have surprising implications. In particular, because it runs it cannot be set to zero for all scales at once. When $\alpha \ne 0$ this can be done for at most a single scale, $\epsilon_0$. This is because $\lambda = 0$ corresponds to $\hat \lambda = d - 2$, which is {\em not} a fixed point of the RG flow unless $\zeta_s = d - 2$ (and so $\alpha = 0$). If there should exist a scale, $\epsilon_0$, for which $\lambda_0 = 0$ then at this scale $\hat \lambda_0 = d - 2$. Because $\zeta_s < d - 2$ this satisfies\footnote{Although this naively satisfies the bound-state condition, the size of this state satisfies $\epsilon_\star \ll \epsilon_0$ and so the state does not lie believably within the low-energy theory.} $|\hat \lambda_0| > \zeta_s$ and so the RG flow is on a trajectory that flows down from $\hat \lambda = + \infty$ at $\epsilon = \epsilon_\star$, passing through $\hat \lambda = d-2$ at $\epsilon = \epsilon_0$ and continuing to fall until asymptotically reaching $\hat \lambda = + \zeta_s < d-2$ as $\epsilon \to \infty$. 

Of later interest is the asymptotic form for this running as $\hat\lambda(\epsilon)$ nears the fixed points at $\hat\lambda = \pm\zeta_s$. Using the asymptotic expression for $\tanh z$ for large positive or negative $z$ leads to 
\bea   \label{lamapp}
 \frac{\hat \lambda(\epsilon)}{\zeta_s}   &=& \frac{ (\hat \lambda_0/\zeta_s) + \tanh\left[ \frac12 \, \zeta_s \ln(\epsilon/\epsilon_0) \right]}{1 + (\hat \lambda_0/\zeta_s) \, \tanh\left[ \frac12 \, \zeta_s \ln(\epsilon/\epsilon_0)\right]} \nn\\ 
 &\simeq&   1 +  2\left( \frac{\epsilon_0}{\epsilon} \right)^{\zeta_s} \left( \frac{\hat\lambda_0 -\zeta_s}{\hat\lambda_0 +\zeta_s} \right) + \cO\left[ \left( \frac{\epsilon_0}{\epsilon} \right)^{2\zeta_s} \right]   \quad\quad\; \hbox{(for $\epsilon \gg \epsilon_0$)}  \\ 
 &\simeq&   -1 -  2\left( \frac{\epsilon}{\epsilon_0} \right)^{\zeta_s} 
 \left( \frac{\hat\lambda_0 +\zeta_s}{\hat\lambda_0 -\zeta_s} \right) + \cO\left[ \left( \frac{\epsilon}{\epsilon_0} \right)^{2\zeta_s} \right]     \quad\quad\; \hbox{(for $\epsilon \ll \epsilon_0$)} \,.\nn
\eea
which reveals how the quantity $d - 2-\zeta_s = d - 2-\sqrt{(d - 2)^2 - 4\alpha}$ acts as an `anomalous dimension' for $\hat\lambda$. In particular if $\hat \lambda_0 = d - 2$ (as appropriate if $\lambda$ vanishes at $\epsilon_0$) the approach to large $\epsilon$ is given by
\be
 \frac{\hat \lambda(\epsilon)}{\zeta_s}  \simeq    1 +  2\left( \frac{\epsilon_0}{\epsilon} \right)^{\zeta_s} \left( \frac{d - 2-\zeta_s}{d - 2+\zeta_s} \right) + \cdots \quad\quad\; \hbox{(for $\epsilon \gg \epsilon_0$ when $\hat\lambda_0 = d - 2$)} \,.
\ee

\section{Scattering and Catalysis}
\label{sec:apps}

We next describe several applications of the above boundary conditions to scattering problems. Doing so illustrates several separate points. First, it shows that the same renormalizations required to make sense of the bound-state problem also do so automatically for scattering problems. Second, it shows how scattering cross sections need not depend on the coupling $\lambda$ in the naive way expected from the Born approximation, $\sigma \propto |\lambda|^2$, which might be expected to vanish --- or be suppressed by powers of the UV scale $(k \epsilon)^p$ --- in the limit where $k\epsilon \to 0$. Instead we find them to be controlled by the RG-invariant scale $k \epsilon_\star$ or $k \epsilon_0$, which can be much larger than the UV scale $\epsilon$ associated with the underlying structure of the source. 

For concreteness we restrict ourselves here to scattering in three dimensions, and specialize at the end to a particularly simple example (non-relativistic $s$-wave scattering of fermions from a magnetic monopole). As a reality check, in an Appendix we also present results for scattering in one dimension since this allows us to connect our framework to standard results for that particularly simple case. 

\subsection{Scattering when $d=3$}

For scattering problems the radial part of the Schr\"odinger equation in three dimensions becomes
\be \label{SchEgksc}
   r^2 \frac{\exd^2 \psi}{\exd r^2} + 2r \, \frac{\exd \psi}{\exd r} + \left( v + k^2 r^2 \right) \psi
 = 0\,,
\ee
where $v = \alpha - \ell(\ell+1)$ since $\varpi = \ell(\ell + 1)$ in three dimensions. The two linearly independent radial profiles that solve this are given in terms of confluent hypergeometric functions by
\be
 \psi_\pm(r)  = (2ik r)^{\frac12(-1\pm \zeta)} \, e^{-ik r}  \cM\left[ \frac12 \left(1 \pm \zeta\right), 1 \pm \zeta ; 2ik r \right] \,,
\ee
and as before, but this time specialized to $d=3$, $\zeta := \sqrt{1 - 4v} = \sqrt{(2\ell+1)^2 - 4 \alpha} = 2l+1$. Writing the general solution as 
\be \label{pmformsc}
 \psi_\infty(r) = C_+ \, \psi_+(r) + C_- \, \psi_-(r)   \,,
\ee
we may in this case use the boundary condition \pref{Rexp} through \pref{Rvslamzet} at the origin to determine the ratio $C_-/C_+$, leading to $C_- = 0$ for $\ell \ne 0$ (for simplicity we restrict, as above, to $0 \le \alpha < \frac54$) and to the expression
\be \label{ratio}
  \frac{C_-}{C_+} = \left(2ik\epsilon \right)^{\zeta_s} \frac{\zeta_s - \hat\lambda}{\zeta_s + \hat\lambda} \,. 
\ee
for $\ell = 0$. Notice that this $\ell =0$ result reduces to $C_- = 0$ if $\hat\lambda$ should lie at the IR fixed point $\hat\lambda = \zeta_s$.

It is the running of $\lambda$ in this expression that makes the ratio $C_-/C_+$ $\epsilon$-independent, and it is the {\em difference} between $\hat \lambda$ and its IR fixed point (rather than, say, having $\lambda = 0$) that drives its value away from the value found in the absence of a source. It is also this running that ensures the size of $C_-/C_+$ ends up being controlled by RG-invariant scales (that can be surprisingly different than the size of the underlying source). 

To see how this works it is convenient to exploit the $\epsilon$-independence of the result to evaluate it asymptotically close to the IR fixed point at $\hat \lambda = \zeta_s$ using \pref{lamapp}. Inserted into \pref{ratio} this shows how the explicit powers of $\epsilon$ cancel to leave
\be \label{ratioapprox}
  \frac{C_-}{C_+} = \left(2ik\epsilon \right)^{\zeta_s} \frac{\zeta_s - \hat\lambda}{\zeta_s + \hat\lambda} \simeq \left(2ik\epsilon_0 \right)^{\zeta_s}   \frac{\zeta_s - \hat\lambda_0}{\zeta_s + \hat\lambda_0}  = -y (2i k \epsilon_\star)^{\zeta_s} \,,
\ee
where the last equality is expressed in terms of an RG-invariant scale $\epsilon_\star$, defined by $\hat\lambda(\epsilon_\star) = \infty$ (if $|\hat\lambda_0| > \zeta_s$) or $\hat \lambda(\epsilon_\star) = 0$ (if $|\hat \lambda_0| < \zeta_s$). Here 
\be \label{ydef}
  y = \hbox{sign}[|\hat \lambda|-\zeta_s]
\ee
 is the RG-invariant sign that determines which definition of $\epsilon_\star$ is to be used.

Finally the ratio $C_-/C_+$ may be related to the scattering amplitude by matching the large-$r$ behaviour of our wavefunction to the form
\be
\label{LL1}
\psi \approx A_\ell\frac{e^{i(kr - \ell\pi/2)}} r + B_\ell\frac{e^{-i(kr - \ell\pi/2)}} r,
\ee
from which we determine the phase shift by \cite{LL}
\be
\label{LL2}
e^{2i\delta_\ell} = - \frac{A_\ell} {B_\ell}.
\ee

To use this we write the asymptotic behaviour of the $\cM$ functions for large imaginary argument as
\be
\label{asymM}
\cM(a, b; z) \approx \frac{\Gamma(b)}{\Gamma(b-a)}e^{a\pi i} z^{-a} + \frac{\Gamma(b)}{\Gamma(a)}e^z z^{-(b-a)}, 
\ee
to find
\bea
\label{andSo}
\psi_\pm &\propto & \frac{\Gamma(1\pm \zeta)}{\Gamma\left[\frac12\left( 1 \pm \zeta\right)\right]}  \left( \frac{ 1}{2ikr} \right)  \Bigl[ e^{ikr} +    e^{\frac12 \left( 1 \pm \zeta \right) i\pi} \;  e^{-ikr}   \Bigr] \,,   \nn\\
&&\quad {}
\eea
and so
\bea
A_\ell &\propto& \left[\frac{\Gamma(1 + \zeta)}{\Gamma\left[\frac12\left( 1 + \zeta\right)\right]} + \left( \frac{C_-}{C_+} \right) \frac{\Gamma(1 - \zeta)}{\Gamma\left[\frac12\left( 1 - \zeta\right)\right]} \right] e^{i\pi \ell/2}, \,\, \text{and} \label{coefA} \\
B_\ell &\propto& \left[\frac{\Gamma(1 + \zeta)}{\Gamma\left[\frac12\left( 1 + \zeta\right) \right]} + \left( \frac{C_-}{C_+} \right) \frac{\Gamma(1 - \zeta)}{\Gamma\left[\frac12\left( 1 - \zeta\right) \right]} \; e^{-i \pi \zeta} \right] e^{(  1 + \zeta - \ell)i\pi /2} \,. \label{coefB}
\eea

For any $\ell \ne 0$ this simplifies using $C_- = 0$ to become 
\be
\label{phaseL}
  \delta_\ell = \frac{\pi}4 \left( 1 - \zeta + 2 \ell \right) \quad \hbox{(for $\ell \ne 0$)} \,.
\ee
For $\ell = 0$, Euler's duplication formula in the form $\Gamma(1\pm \zeta)/\Gamma\left[ \frac12 \left(1 \pm \zeta\right) \right] = 2^{\pm \zeta} \Gamma\left(1 \pm \frac12 \, \zeta \right)/\sqrt\pi$, instead gives the phase shift  
\be
\label{phase}
e^{2i\delta_0} = \left[ \frac{1 + \cA \,e^{i\pi \zeta_s/2}}{1 +\cA \,e^{-i \pi \zeta_s/2} } \right] e^{(1 - \zeta_s)i\pi /2} \,,
\ee
where (with $y$ as defined in \pref{ydef})
\be
  \cA := -y \left( \frac{k \epsilon_\star}{2} \right)^{\zeta_s} \left[ \frac{\Gamma\left(1 - \frac12 \, \zeta_s \right)}{ \Gamma\left(1 + \frac12 \, \zeta_s \right) } \right] \,.
\ee

Notice that this depends on the delta-function coupling only through the RG-invariant quantity $y \epsilon_\star$. It is this feature in particular that opens the door to the possibility of catalysis: contact interactions can contribute to observables (such as scattering rates) by an amount much larger than the UV size naively associated with the microscopic source. Such catalysis occurs in situations where RG evolution predicts values for $\epsilon_\star$ to be much larger than the values appearing in $\hat\lambda(\epsilon)$ and $\epsilon$:
\be
 \epsilon_\star = \epsilon \, \exp \left[ - \frac{2}{\zeta_s} \tanh^{-1} \left( \frac{\zeta_s}{\hat\lambda} \right) \right] \,. 
\ee
This gives $\epsilon_\star \gg \epsilon$ when $\hat\lambda(\epsilon)$ is close to the UV fixed point: $\hat\lambda(\epsilon) \simeq - \zeta_s(1 + \delta)$, where $|\delta| \ll 1$. 

\subsection*{Special case: the delta-function potential}

Setting $\alpha = 0$ in the above formulae reduces them to a useful special case: where the scattering is purely from the delta-function part of the interaction. In this limit \eqref{phaseL} and \eqref{phase} give the phase shift for scattering from the 3D delta-function potential when evaluated with $\zeta = 2\ell + 1$. In this limit the point $\lambda = 0$ (and so $\hat\lambda = 1$) becomes a fixed point of the RG flow, at which point there is no scattering. Away from this fixed point \eqref{phaseL} shows that $\delta_\ell = 0$ for $\ell \neq 0$, so only $s$-wave scattering is nontrivial. 

To compute the $s$-wave phase shift, substituting $\zeta_s = 1$ into \eqref{phase} gives
\be
\label{deltaPhase}
e^{2i\delta_0} = \frac{1 + i\cA}{1 - i\cA}, 
\ee
with
\be
\label{deltaA}
\cA = -y k\epsilon_\star \,.
\ee
Equivalently
\be
\label{tanPhase}
\tan \delta_0 = \frac{\Im(1 - iy k\epsilon_\star)}{\Re(1 - iy k\epsilon_\star)} = -y k\epsilon_\star \,,
\ee
in agreement with standard calculations \cite{Jackiw}. The energy dependence of this result is captured by the scattering length, $a_s$, since $a_s = -\frac{1}{k}\tan \delta_0 = y \epsilon_\star$ is $k$-independent. This comparison gives an independent measure of the RG-invariant scale $\epsilon_\star$, and using this to trade $y\epsilon_\star$ for $a_s$ in the earlier predictions for bound-state energy shifts provides them as functions of $a_s$; thereby directly relating physical observables to one another. As applied to mesic atoms (such as a $\pi^-$ or $K^-$ orbiting a nucleus) \cite{KG} similar reasoning leads to the Deser formula \cite{Deser} relating energy-level shifts due to nuclear forces to nuclear scattering lengths.

\subsection{Nonrelativistic $s$-wave fermion-monopole scattering}

The delta-function scattering result just described has an immediate application to $s$-wave elastic scattering of non-relativistic fermions from a magnetic monopole, and shows how these processes might also exhibit catalysis. In this section we develop this connection a bit more explicitly, with the goal of showing that this type of scattering can also be controlled by the RG-invariant scale defined by the contact interaction.\footnote{Whether or not this leads to catalysis depends on whether or not matching to the underlying monopole actually does give couplings $\lambda(\epsilon)$ for small $\epsilon$ that lie sufficiently close to the UV fixed point.} 

The utility of something as simple as delta-function scattering to something as complicated as fermion-monopole scattering arises because of the great simplicity of $s$-wave scattering for these systems; in particular the possibility of there being no angular momentum barrier. For scattering from a magnetic monopole the absence of such a barrier for $s$-wave scattering is a bit more subtle than it looks, and would not be possible for monopole scattering of a spinless boson. 

Finally we apply the above insights to non-relativistic $s$-wave scattering of a charged particle from a magnetic monopole in the Pauli approximation of spin, where the running of $\lambda$ provides a simple understanding of why such scattering need not be suppressed by microscopic ({\em i.e.} GUT-scale) lengths, as would have been naively expected in Born approximation. This bears more than passing resemblance to earlier discussions of monopole catalysis of baryon number violation, as we explore in more detail in a companion paper \cite{Dirac} dedicated to the full relativistic treatment. 

The electromagnetic potential of a point-like monopole with magnetic charge $g$ sitting at the origin is given in spherical polar coordinates by 
\be
  \bfA = \frac{g}{4\pi r} \; \tan \frac{\theta}{2} \; \bfe_\phi \,,
\ee
where $\bfe_\phi$ is the unit vector in the azimuthal direction and the Dirac quantization condition enforces $eg = 2\pi n$ for integer $n$, where $e$ is the charge of a particle that moves in this monopole field. As discussed in many references --- see for example the reviews \cite{MonopoleReview} --- the magnetic field contributes to the angular momentum for motion in this potential, leading to a modified form for the conserved orbital angular momentum, $\bfL \ne \bfr \times \bfp$. The eigenvalues of $\bfL^2$ for this modified angular momentum remain $\ell (\ell + 1)$ but now with $\ell = \mu, \mu+1, \mu+2 \,, \cdots$ with $\mu := eg/4\pi = n/2$. 

Consequently $\ell = 0$ is no longer an option for spinless particles scattering from a magnetic monopole. When $\mu = \pm \frac12$ it {\em is} possible for spin-half particles to have zero total angular momentum, however, and so it is for this specific combination that one might hope to find no angular-momentum barrier in the radial equation. This is borne out by detailed calculations, for which the Schr\"odinger-Pauli equation 
\be
\label{monAction}
 \left[ i\partial_t + \frac{1}{2m} \left( \bm{\sigma} \cdot \left[ \bm{\nabla} - ie\mathbf{A}\right]\right)^2 \right] \Psi = 0 \,,
\ee
acting on the 2-component Pauli spinor, $\Psi$, leads to the following radial equation
\be
\label{MonEqa}
\frac{1}{r^2} \partial_r \Bigl( r^2 \, \partial_r \psi \Bigr) - \frac{1}{r^2}\left({\bm{L}}^2 - \mu^2\right)\psi + \frac{\mu}{r^2}\left(\bm{\sigma}\cdot\bfe_r \right)\psi + k^2\psi = 0
\ee
for energy eigenstates $\Psi = \psi e^{-iEt}$ with momentum $k = \sqrt{2mE}$ as above. 

The $\bm{\sigma} \cdot\bfe_r$ term here has its origin in the $\bm{\sigma} \cdot \bfB$ interaction with the fermion's magnetic moment. For $\ell = \mu$ precisely one of the spinor harmonics satisfies $\bm{\sigma} \cdot \hat \bfe_r  \psi = -\psi$, so that the coefficient of $1/r^2$ becomes proportional to $\bfL^2 - \mu^2 - \mu = \mu(\mu+1) - \mu^2 - \mu = 0$, as claimed. In this case the bulk potential interaction vanishes, leaving only the delta-function coupling coming from any contact interaction, such as
\be
\label{LbMon}
\cL_b = -h \Psi^\dagger\Psi \,,
\ee
at the position of the source monopole. This is the mode believed to participate in monopole-catalyzed baryon violation \cite{monopolecatal}, for which the absence of an angular-momentum barrier allows an incoming $s$-wave state to penetrate right down to the monopole position.

For the present purposes what is important is that the scattering story for this mode is told precisely as above, dropping the inverse-square potential and keeping only the delta-function interaction. Consequently its size is controlled by the RG-invariant scale $\epsilon_\star$, raising the possibility that monopole catalysis might be regarded as a special case of the more general catalysis phenomenon discussed above. 

We extend the above discussion to a more detailed examination of the implications of PPEFTs to monopole scattering and to the problem of scattering in the Coulomb potential in several companion papers \cite{KG, Dirac, ProtonR}.

\subsection*{Summary}

In short, we propose the systematic use of the effective point-particle action for deriving the boundary conditions appropriate near any source, and show why this point of view helps understand the subtleties of the quantum mechanics of singular potentials like the inverse-square potential. The presence of an action makes transparent the process of regularization and renormalization of fields that are classically singular at the source, and allows use of renormalization-group techniques for resumming the breakdown of perturbation theory that often arises in the near-source limit due to these singularities.

The great virtue of this approach is its constructive nature, since the linking of boundary conditions to source properties removes the guesswork associated with their determination. Furthermore, because only a few types of actions typically dominate at low energies (for instance a choice quadratic in fields often does, as we describe here) it is likely that a very broad variety of systems will fall onto a universal class of renormalization-group flows.

\section*{Acknowledgements}

We thank Horacio Camblong, Friederike Metz, Duncan O'Dell, Carlos Ordon\~ez, Ryan Plestid and Markus Rummel for useful discussions and Ross Diener, Leo van Nierop and Claudia de Rham for their help in understanding singular fields and classical renormalization. CP${}^3$ Odense and the Neils Bohr Institute kindly hosted CB while part of this work was done. This research was supported in part by funds from the Natural Sciences and Engineering Research Council (NSERC) of Canada, by a postdoctoral felloship from the National Science Foundation of Belgium (FWO) and by the Ontario Trillium Fellowship. Research at the Perimeter Institute is supported in part by the Government of Canada through Industry Canada, and by the Province of Ontario through the Ministry of Research and Information (MRI).  

\appendix

\section{The boundary action and the meaning of the classical RG}
\label{AppBC}

This appendix provides a more systematic derivation of the boundary conditions used, following closely the reasoning used in \cite{EFTCod2}. More careful reasoning is required because the standard argument becomes suspicious in the presence of a sufficiently singular potential. The argument of \cite{EFTCod2} has the added value of providing a simple interpretation for what the classical RG equations physically mean.

\subsection*{The delta-function story}

Recall first the argument establishing the boundary conditions used in the main text. This uses the naive reasoning appropriate for delta-function interactions in the absence of singularities in the rest of the scalar potential. In it one starts with the time-independent Schr\"odinger equation (say),
\be 
 - \nabla^2 \psi +\Bigl[  U(\bfr) + \lambda \, \delta^3(\bfr) \Bigr] \psi = k^2 \psi \,,
\ee
and integrates over the small pillbox, $\cS$, with radius $\epsilon$ centred at $\bfr = 0$. Now comes the main point: {\em if} $U(\bfr)$ is smooth within $\cS$, then one expects the integral of $[U(\bfr) - k^2] \psi$ over $\cS$ to vanish in the limit of vanishing pillbox radius, $\epsilon \to 0$. When this is true the remaining factors give
\be \label{appbc1}
 4\pi \epsilon^2 \, \frac{\partial \psi}{\partial r} = \int_\cS \exd^3 x \; \nabla^2 \psi = h \, \psi(0) \,,
\ee
leading to the boundary condition used in the main text. However if $U(\bfr)$ is sufficiently singular at $r = 0$ its integral over $\cS$ need not vanish as required when $\epsilon \to 0$, undermining faith in its validity.

\subsection*{A better argument}

To arrive at a better argument it helps to refer again to Fig.~\ref{Figcirc}, which identifies three scales subject to the hierarchy $\varepsilon \ll \epsilon \ll a$, where $\varepsilon$ and $a$ are physical scales respectively associated with the size of the underlying source and the size of the physics that is of interest far from the source. The scale $\epsilon$ is not part of the original physical problem, however, but rather is introduced purely as a calculational crutch in order to conceptually separate the calculation into two parts. One imagines drawing a sphere, $\cS$, of radius $\epsilon$ around the source and then separately thinking about the large-distance story relevant for $r \gg \epsilon$ outside this sphere and about a small-distance story relevant to $r \ll \epsilon$ well inside this sphere.

The two steps of the problem are then to derive the fields and their derivatives at the surface of $\cS$ given the properties of the source (typically as specified by its action, $S_b$). The advantage is that this can be done once and for all, without specifying precisely which observables are of interest outside $\cS$. These enter only in the second step, where their properties are computed using only the boundary data on the surface of $\cS$. 

Although in principle $\epsilon$ could be arbitrary, in practice we choose $\epsilon \gg \varepsilon$ in order to profit from a multipole expansion (or its generalization) when computing the boundary information on the surface of $\cS$, since successive multipoles are suppressed by powers of $\varepsilon/\epsilon$. We similarly choose $a \gg \epsilon$ so that the observables of interest are sufficiently distant from $\cS$ that they are not inordinately sensitive to boundary effects there. 

\subsection*{The boundary action}

Literally specifying fields and their radial derivatives at the boundary, $\cB = \partial\cS$, is overkill, however, since in general it would over-determine the boundary-value problem exterior to $\cS$. This is because in practice the actual values taken by fields on $\cB$ also depend somewhat on the positions of other source and boundaries elsewhere in the problem, and although this dependence is weaker the further away they are (hence the condition $\epsilon \ll a$), it is there and the boundary information at $\cB$ must be encoded in a way that leaves the fields free to adjust as required to meet its demands.
 
A simple and efficient way to do so is to specify the boundary data at $\cB$ in terms of a {\em boundary action}, $I_\cB$. This action is related to, but not the same as, the original source action, $S_b$. Whereas $S_b$ is an integral over the source's world-volume, $I_\cB$ is always integrated over a codimension-1 world-volume that $\cB$ sweeps out as time evolves. For instance for point particles in 3 spatial dimensions $S_b$ comes to us as a one-dimensional integral over the particle world-line, while $I_\cB$ is a 3-dimensional integral over time and the two angular directions of a 2-sphere surrounding the particle. For instance, for the type of interaction studied in the main text we have (using adapted coordinates in the particle rest frame)
\be \label{SvsI}
  S_b = - \int \exd t \, h \, \psi^* \psi \quad \hbox{and} \quad
  I_\cB = - \int \exd^3x  \, \tilde h \, \psi^*\psi \,,
\ee
and so whereas $h$ has dimension length${}^2$ --- for the 3D Schr\"odinger field, for which $\psi^* \psi$ is length${}^{-3}$ --- the coupling $\tilde h$ is dimensionless.

Once $I_\cB$ is specified the surface $\cB$ can be regarded as a boundary of the exterior region, with its influence on physics exterior to $\cS$ obtained in the usual way.\footnote{That is, the source physics contributes to the path integral a phase $e^{iI_\cB}$, or phrased another way, classical physics is determined by requiring the total action, $S_\ssB + I_\cB$, be stationary with respect to variations of the fields throughout the bulk {\em and} on the boundary surface $\cB$.} 

What is needed is a way to construct $I_\cB$ given $S_b$. In principle this is a matching calculation: if $N$ effective couplings are of interest on $S_b$ then we compute any $N$ convenient quantities exterior to $\cS$ from which they could be determined. We then compute these same $N$ quantities in terms of the most relevant interactions on $I_\cB$ and equate the results in order to infer the couplings of $I_\cB$ in terms of those of $S_b$. 

For the simple cases considered in the main text our interest is largely restricted to a single term in the source action, like $h\, \psi^*\psi$, and so this matching process is particularly easy. In practice in such situations the relation between $S_b$ and $I_\cB$ boils down to something very simple: $S_b$ is the dimensional reduction of $I_\cB$ (as would be appropriate if $\cS$ were examined with insufficient resolution to see that it has a finite radius). For instance the couplings of \pref{SvsI} would be related by
\be \label{hhtrel}
  \tilde h = \frac{h}{4\pi \epsilon^2 } \,.
\ee

Finally, given the boundary action $I_\cB$ of \pref{SvsI} the boundary condition is simple to derive by varying the combined bulk+boundary action $S_\ssB +I_\cB$ with respect to variations of the fields on the boundary $\cB$. In the present instance only the spatial derivatives of $S_\ssB$ play any role, due to the integration by parts required when computing $\delta S_\ssB/\delta \psi^*$, generalizing the scalar boundary condition \pref{appbc1} to
\be \label{appbc2} 
\left[  \frac{\delta S_\ssB}{\delta \psi^*} + \frac{\delta I_\cB}{\delta \psi^*} \right]_{r = \epsilon} = \left[ \frac{\partial \psi}{\partial r} + \frac{\delta I_\cB}{\delta \psi^*} \right]_{r = \epsilon} = 0 \,,
\ee
where the first equality assumes $S_\ssB = - \int \exd^4 x [ \nabla \psi^* \cdot \nabla \psi + \cdots ]$, where none of the terms represented by ellipses require integration by parts\footnote{The variation of the terms in the ellipses are themselves cancelled by the rest of $\delta S_\ssB/\delta \psi^*$, inasmuch as they appear in the $r \to \epsilon$ limit of the bulk field equation.} when evaluating $\delta S_\ssB/\delta \psi^*$. In situations where $S_b = 4\pi \epsilon^2 I_\cB$ this takes the form found in \cite{EFTCod2,BDg}:
\be \label{appbc3} 
 4\pi \epsilon^2 \; \left( \frac{\partial \psi}{\partial r} \right)_{r = \epsilon} = -\frac{\delta S_b}{\delta \psi^*}  \,. 
\ee
This is the boundary condition used in the main text, and shows how it would be generalized to situations beyond the simplest $\psi^* \psi$ interaction. 

What is clear is that this construction goes through equally well with and without there being an inverse-square potential in the bulk, and this is the underlying reason why the singularity of the inverse-square potential does not alter the boundary condition used in the main text.

\subsection*{Physical interpretation of the RG}

The boundary action provides a simple physical interpretation of what the RG evolution described in the text. What it expresses is that there is nothing unique about the choice of $\cS$, whose radius, $\epsilon$, could be anything subject only to the condition $\varepsilon \ll \epsilon \ll a$. The RG condition expresses this fact.

In detail, when computing the RG evolution by differentiating with respect to $\epsilon$ we always hold fixed all physical quantities. What we are doing is adjusting the couplings in $I_\cB$ in such a way as to not change the physical bulk-field profile. This relates the RG evolution of the couplings in $I_\cB$ to the classical bulk field equations.\footnote{This resembles what occurs in AdS/CFT \cite{ADSCFT}} The couplings of $S_b$ then inherit this because of relations like \pref{hhtrel}. This is why it is the combination $h/(4\pi\epsilon^2)$ that naturally arose in the RG derivation given in the main text.

\section{Scattering in 1 dimension}

This appendix computes the transmission and reflection coefficients implied by the main text in the special case of scattering in $d = 1$ space dimension. Besides providing a comparison with standard results, it also illustrates (as in the 3D case) that the source renormalization required to make sense of bound-state calculations automatically does the same for other observables, and shows how the RG-invariant scale can be related to  scattering observables.

For scattering we take $E > 0$ and so $\kappa = ik$ is imaginary with $k^2 = 2mE \ge 0$. Eq.~\pref{SchE} becomes
\be \label{SchEgSc}
  \partial_x^2 \psi + \frac{\alpha}{x^2} \, \psi = -k^2 \, \psi  \,,
\ee
where $\alpha = 2m g$ as before. The change of variables $\psi(x) = \sqrt{z} \; u(z)$ with $z = k x$ implies $u(z)$ satisfies Bessel's equation,
\be \label{mBESc}
 z^2 \, u'' + z \, u' +(z^2 - \sigma^2) u = 0 \,,
\ee
with $\sigma^2 := \frac14 - \alpha = 4\zeta^2$. 

For $E \ge 0$ eq.~\pref{mBESc} is most usefully solved by the Hankel functions $H^{(1)}_\sigma(z)$ and $H^{(2)}_{\sigma}(z)$, which in our conventions (see next Appendix) asymptote for large real $z$ to
\bea
 H_\sigma^{(1)}(z) &\simeq& \sqrt{\frac{2}{\pi z}} \; \exp\left[ i \left( z - \frac{\pi \sigma}{2} - \frac{\pi}4 \right)  \right]  \nn\\
 \hbox{and} \quad
 H_\sigma^{(2)}(z) &\simeq& \sqrt{\frac{2}{\pi z}} \;  \exp\left[ -i \left( z - \frac{\pi \sigma}{2} - \frac{\pi}4 \right) \right] \,.
\eea
Since $H_\sigma^{(1)}(kx) \, e^{-ik t}$ asymptotes at large $x$ to be proportional to $e^{-ik(t-x)}$ it represents a right-moving wave, whilst $H_\sigma^{(2)}(kx)$ similarly asymptotes at large $x$ to a left-moving wave. Similar expressions hold for negative $x$, as may be found using the  reflection properties of the Hankel functions given in the Appendix. 

For a particle that initially approaches from $x \to +\infty$ we therefore take the following solutions for $x > \epsilon$ and $x < -\epsilon$:
\bea
 \psi_+(x) &=& \cN \sqrt{kx} \Bigl[ H_{\sigma}^{(2)}(kx) + \cR \, H_{\sigma}^{(1)}(kx)\Bigr] \quad \hbox{for $x \ge \epsilon$}\nn\\
 \hbox{and} \quad 
 \psi_-(x) &=& \cN \sqrt{kx} \; \cT \, H_{\sigma}^{(2)} (kx)  \qquad\qquad\qquad \hbox{for $x \le -\epsilon$}\,,
\eea
where $\cN$ is an irrelevant constant and the reflection and transmission coefficients, $\cR$ and $\cT$, are the integration constants to be determined using the boundary condition at $x = \pm \epsilon$. 

Demanding $\psi_+(\epsilon) = \psi_-(-\epsilon)$ and using the reflection properties of the Appendix implies
\be
 H_{\sigma}^{(2)}(k\epsilon) + \cR \, H_{\sigma}^{(1)}(k\epsilon)  = i e^{i\pi \sigma} \cT \, H_{\sigma}^{(1)} (k\epsilon) \,,
\ee
and so using the small-$k\epsilon$ limit of the Hankel functions then gives,
\be \label{conteq}
 \cR - i\cT \, e^{i\pi\sigma}  = -\,  \frac{ H_{i\nu}^{(2)}(k\epsilon) }{ H_{i\nu}^{(1)} (k\epsilon)} =- \,  \frac{-1+X\, e^{i\pi\sigma}}{1-X \, e^{-i\pi\sigma}} = \frac{1- X\, e^{i\pi\sigma}}{1-X \, e^{-i\pi\sigma}}  \,,
\ee
where 
\be
 X := \frac{\Gamma(1-\sigma)}{\Gamma(1+\sigma)} \left( \frac{k\epsilon}{2} \right)^{2\sigma} \,.
 \ee
The jump condition relating the derivatives at $x = \epsilon$ to $\lambda$ gives a second relation between $\cR$ and $\cT$
\bea
 \lambda &=& \partial_x \ln \psi_+(\epsilon) - \partial_x \ln \psi_-(- \epsilon) \nn\\
 &=&  \frac{1}{\epsilon} \left\{ 1 - \sigma\left[ \frac{ (1+ X\, e^{i\pi\sigma}) - \cR (1+ X\, e^{-i\pi\sigma}) }{1-X \, e^{i\pi\sigma} - \cR \left(1 - X\, e^{-i\pi\sigma} \right)} + \frac{  1+X\, e^{-i\pi\sigma}}{1 - X\, e^{-i\pi\sigma} } \right]\right\} \,.
\eea
The divergence here as $\epsilon \to 0$ is, as before, cancelled by the $\epsilon$-dependence implicit in $\lambda$, and can in principle be determined by differentiating with respect to $\epsilon$ with $k$ held fixed. But because this boundary condition is precisely the same one that was used to determine this dependence in the bound-state problem this argument is guaranteed to return the same evolution equation for $\lambda$ that was found earlier.

For the sake of concreteness we specialize now to $\sigma > 0$ (and so $\alpha < \frac14$) and $|\hat \lambda| > 2\sigma$. To write a more explicit form for $\cR$ and $\cT$ it is useful to exploit the RG-invariance of the above conditions for $\cT$ and $\cR$ by evaluating them using a convenient value for $\epsilon$, which we choose as $\epsilon \ll \epsilon_\star$. This choice allows two simplifications to be used. First, we can use $k \epsilon \ll 1$ to expand in powers of $X$ because $X$ is proportional to a positive power of the small quantity $k\epsilon \ll 1$, leading to
\be  \label{smallX}
  \frac{\hat\lambda }{2\sigma} =   - \left[ 1 + X e^{i\pi\sigma} \left( \frac{1- \cR \, e^{-2\pi i \sigma}}{1 - \cR } \right) + X e^{-i\pi \sigma} +\cO(X^2)\right] \,.\ee
  Second we can use the RG evolution to infer that $\hat\lambda \to \hat\lambda_- = - 2\sigma$ as $\epsilon \to 0$, and so use the asymptotic small-$\epsilon$ expression \eqref{lamapp} for the approach of $\hat\lambda$ to this limit:
\be \label{asymRG}
  \frac{\hat \lambda }{2\sigma} = - \coth \left[ \sigma \ln \left( \frac{\epsilon_\star}{\epsilon} \right) \right] \simeq -1 - 2 \left( \frac{\epsilon}{\epsilon_\star} \right)^{2\sigma} \,.
\ee
Inserting this into \pref{smallX} gives
\bea
  2 \left( \frac{\epsilon}{\epsilon_\star} \right)^{2\sigma} &=& X \left[ e^{i\pi\sigma} \left( \frac{1}{1 - \cR } \right) +  e^{-i\pi \sigma}  \left(1- \frac{\cR }{1 - \cR } \right) \right]  \nn\\
  &=& \left( \frac{k \epsilon}{2} \right)^{2\sigma} \frac{\Gamma(1-\sigma)}{\Gamma(1+\sigma)} \left[ e^{i\pi\sigma} \left( \frac{1}{1 - \cR } \right) +  e^{-i\pi \sigma}  \left(1- \frac{\cR }{1 - \cR } \right) \right]   \,,
\eea
from which the $\epsilon$-dependence completely drops out, as it must. 

The reflection coefficient, $\cR$, found by solving this last equation then is
\be
 \cR = \frac{ X_\star \cos(\pi\sigma) - 1}{X_\star \, e^{-i\pi\sigma}  - 1} \,,
\ee
where $X_\star = X(\epsilon = \epsilon_\star)$. Using this in the small-$k\epsilon$ limit of \pref{conteq} then implies 
\be
 \cT = i e^{-i\pi\sigma}(1 - \cR)=  \frac{ X_\star \, e^{-i\pi\sigma}   \sin(\pi\sigma) }{X_\star \, e^{-i\pi\sigma}  - 1}  \,.
\ee
Notice these satisfy $|\cR|^2 + |\cT|^2 = 1$ when $X_\star$ is real, and reduce to the appropriate limit of delta-function scattering in the limit $\alpha \to 0$ (and so $\sigma \to \frac12$), which are
\be
  \cR  \to  \frac{1}{1- 2i  k /  \lambda_-  } = \frac{1}{1 + ik \epsilon_\star}  \quad \hbox{and} \quad 
    \cT \to  - \frac{2ik/\lambda_-}{1-2ik/\lambda_-} = \frac{ ik \epsilon_\star }{1+ik\epsilon_\star} \quad \hbox{($\alpha \to 0$)}\,.
\ee
These last expressions use $\lambda_- = -2/\epsilon_\star$, corresponding to using $\sigma = \frac12$ in \pref{asymRG} to learn $\hat\lambda = \epsilon \lambda - 1$ asymptotes to $\hat\lambda_- = - 2\sigma = -1$ as $\epsilon \to 0$, and so $\lambda$ itself satisfies
\be 
  \epsilon \lambda_-  \simeq - 2 \left( \frac{\epsilon}{\epsilon_\star} \right) \,,
\ee
in the small-$\epsilon$ regime.
 
The special case where $\sigma \to 0$ is also a delicate one since in this limit
\be
 X_\star := \frac{\Gamma(1-\sigma)}{\Gamma(1+\sigma)} \left( \frac{k\epsilon_\star}{2} \right)^{2\sigma} \to 1 + 2\sigma \left[ \gamma + \ln \left( \frac{k \epsilon_\star}{2} \right)\right] + \cO(\sigma^2) =: 1 + \cA_\star \sigma + \cO(\sigma^2)\,,
\ee
where the last equality defines $\cA_\star$ and $\gamma$ is the Euler-Mascheroni constant.  Consequently $X_\star \, e^{-i\pi\sigma}  - 1 =  (\cA_\star - i \pi) \sigma + \cO(\sigma^2)$, while $X_\star \, e^{-i\pi\sigma}   \sin(\pi\sigma) = \pi \sigma + \cO(\sigma^2)$ and $X_\star \cos(\pi\sigma) - 1=  \cA_\star  \sigma + \cO(\sigma^2)$, leading to
\be
  \cR \to   \frac{\cA_\star}{\cA_\star - i \pi} \quad \hbox{and} \quad \cT \to  \frac{\pi}{\cA_\star - i \pi}\quad \hbox{(when $\sigma \to 0$)} \,.
\ee 

What is noteworthy about these expressions is that the reflection and transmission amplitudes are controlled by the RG-invariant combination $k \epsilon_\star$ rather than by the value of $\lambda(\epsilon)$ as measured at a microscopic scale $\epsilon$. This is particularly striking in the limit $\alpha \to 0$ for which it is only the delta-function potential that drives the scattering. It is in strong contrast to what one might have expected if treating a delta-function potential in Born approximation, since then would have been simply depended on $\lambda$ rather than the fixed-point value $\lambda_-$. This carries the seeds of the explanation of why microscopic scales can drop out of scattering cross sections --- such as is known to be the case for monopole catalysis of baryon-number violation \cite{monopolecatal} ---  and of a more systematic effective description of such processes.

\subsection*{Bessel function properties}

This appendix gathers useful properties of Bessel functions, defined as the series solutions to
\be \label{mBEScApp}
 z^2 \, u'' + z \, u' +(z^2 - \sigma^2) u = 0 \,,
\ee
of the form
\bea
 J_\sigma(z) &=& \sum_{n=0}^\infty \frac{(-)^n}{n! \, \Gamma(n+\sigma+1)} \, \left( \frac{z}{2} \right)^{2n+ \sigma} 
 \nn\\
 &=& \frac{1}{\Gamma(\sigma+1)} \, \left( \frac{z}{2} \right)^{\sigma}\left[1 - \frac{z^2}{4(\sigma+1)} + \cO(z^{4}) \right] \,.
\eea
It is also often useful to use the explicit form of the linearly independent solution
\be
 N_\sigma(z) = \frac{J_\sigma(z) \, \cos(\pi \sigma) - J_{-\sigma}(z)}{\sin (\pi \sigma)} \,.
\ee

For large real $z$ these enjoy the asymptotic forms
\bea
 J_\sigma(z) &=& \sqrt{\frac{2}{\pi z}} \; \left[ \cos \left( z - \frac{\pi \sigma}{2} - \frac{\pi}4 \right) + \cO(1/z) \right] \nn\\
 \hbox{and} \quad
 N_\sigma(z) &=& \sqrt{\frac{2}{\pi z}} \; \left[ \sin \left( z - \frac{\pi \sigma}{2} - \frac{\pi}4 \right) + \cO(1/z) \right] \,.
\eea

For scattering problems of more use are the Hankel functions, defined by
\bea
  H_\sigma^{(1)}(z) = J_\sigma(z) + i N_\sigma(z) = \frac{J_{-\sigma}(z) - e^{-i\pi\sigma} J_\sigma(z)}{i \sin(\pi \sigma)} \nn\\
  \hbox{and} \quad
  H_\sigma^{(2)}(z) = J_\sigma(z) - i N_\sigma(z) = \frac{J_{-\sigma}(z) - e^{i\pi\sigma} J_\sigma(z)}{-i \sin(\pi \sigma)} \,,
\eea
since these satisfy (for large real $z$) 
\bea
 H_\sigma^{(1)}(z) &\simeq& \sqrt{\frac{2}{\pi z}} \; \exp\left[ i \left( z - \frac{\pi \sigma}{2} - \frac{\pi}4 \right)  \right]  \nn\\
 \hbox{and} \quad
 H_\sigma^{(2)}(z) &\simeq& \sqrt{\frac{2}{\pi z}} \;  \exp\left[ -i \left( z - \frac{\pi \sigma}{2} - \frac{\pi}4 \right) \right] \,.
\eea

The property $J_\sigma(-z) = e^{i\pi \sigma} J_\sigma(z)$ shows that on reflection the Hankel functions satisfy
\be
 H_\sigma^{(1)}(e^{\pm i \pi } z) = \frac{J_{-\sigma}(e^{\pm i \pi } z) - e^{-i\pi\sigma} J_\sigma( e^{\pm i \pi } z)}{i \sin(\pi \sigma)} = \frac{e^{\mp i\pi\sigma} J_{-\sigma}(z) - e^{i\pi\sigma(-1\pm1)}J_\sigma(z)}{i \sin(\pi \sigma)} \,,
\ee
and so the direction of the phase rotation makes a difference. In particular, of use in the main text is
\be \label{reflApp}
 H_\sigma^{(1)}(e^{i\pi} z) = - e^{-i\pi \sigma} H_\sigma^{(2)}(z) \quad \hbox{and similarly} \quad H_\sigma^{(2)}(e^{-i\pi }z) = - e^{i\pi\sigma} H_\sigma^{(1)}(z) \,,
\ee
though the formulae are more complicated if we rotate in the other direction by $\pi$. 

For small $z$ we have the expansions
\bea
 H_\sigma^{(1)}(z) &=& \frac{1}{i \sin(\pi \sigma)} \left\{ \frac{1}{\Gamma(1-\sigma)} \left( \frac{z}{2} \right)^{- \sigma} \left[ 1 - \frac{z^2}{4(1-\sigma)} + \cO(z^4) \right] \right. \nn\\
 && \qquad\qquad\qquad\qquad \left. - \frac{e^{-i\pi\sigma}}{\Gamma(1+ \sigma)} \left( \frac{z}{2} \right)^{\sigma}  \left[ 1 - \frac{z^2}{4(1+\sigma)} + \cO(z^4) \right] \right\}   \nn\\
 H_\sigma^{(2)}(z) &=& \frac{1}{i \sin(\pi \sigma)} \left\{ -\, \frac{1}{\Gamma(1-\sigma)} \left( \frac{z}{2} \right)^{- \sigma} \left[ 1 - \frac{z^2}{4(1-\sigma)} + \cO(z^4) \right] \right. \nn\\
 && \qquad\qquad\qquad\qquad \left. + \frac{e^{i\pi\sigma}}{\Gamma(1+ \sigma)} \left( \frac{z}{2} \right)^{\sigma}  \left[ 1 - \frac{z^2}{4(1+\sigma)} + \cO(z^4) \right] \right\}   \,.
\eea

\end{document}